\documentclass[twocolumn,amsmath,amssymb,showpacs]{revtex4}

\usepackage{color}
\usepackage{amsmath}
\usepackage{graphicx}
\usepackage{makecell}

\begin{document}
\title{Simulated scanning tunneling microscopy images of few-layer-phosphorus\\capped by hexagonal boron nitride and graphene monolayers}
\author{Pablo Rivero,$^1$ Cedric M. Horvath$^1$, Zhen Zhu$^2$, Jie Guan$^2$, David Tom{\'a}nek$^{2,*}$, and Salvador Barraza-Lopez$^{1}$}
\email{tomanek@pa.msu.edu, sbarraza@uark.edu}
\affiliation{1. Department of Physics,
             University of Arkansas,
             Fayetteville, AR 72701, USA\\
             2. Physics and Astronomy Department,
             Michigan State University,
             East Lansing, MI 48824, USA}

\begin{abstract}
Elemental phosphorous is believed to have several stable
allotropes that are energetically nearly degenerate, but
chemically reactive. These structures may be capped by
monolayers of hexagonal boron nitride ({\em h}-BN) or graphene to prevent chemical degradation
under ambient conditions. We
perform {\em ab initio} density functional calculations to
simulate scanning tunneling microscopy (STM) images of different
layered allotropes of phosphorus and study the effect of the capping
layers on these images. At scanning energies within its intrinsic conduction gap, protective monolayers of
insulating {\em h}-BN allow to distinguish between the different structural phases of phosphorus underneath due to the electronic hybridization 
with orbitals from the upmost phosphorus atoms: {\em h}-BN capping monolayers
thus provide a promising route to tell few-layer phosphorus allotropes from one another with local probes.
\end{abstract}
\date{\today}

\maketitle
\section{Introduction}
There has been an unprecedented interest in layered phosphorus
allotropes as a new member of the family of two-dimensional (2D)
materials   in the post-graphene era \cite{Novoselov2D,2DACSNano}.
The interest has been triggered by reports indicating that layered bulk black
phosphorus \cite{{Bridgman14},Keyes,Ellis,Jamieson,Schiferl,Morita,%
SSC84,PRB2000,PRL2003} can be exfoliated mechanically yielding
few-layer phosphorene, a novel 2D material with a significant fundamental band gap and
a high carrier mobility \cite{Peide,Yuanbo,Koenig,Buscema} exceeding
that of transition metal dichalcogenides ($MX_2$'s) \cite{ani1}.
Few-layer phosphorene bears promise for intriguing optoelectronics
applications \cite{Jarillo} since the fundamental band gap can be
tuned by the number of layers and by in-layer stretching or
compression \cite{Peide,strain1,strain2,strain3,strain4,strain5,strain6}.
But what truly sets phosphorene apart from other 2D systems is its
polymorphism: Besides the well-studied black phosphorus allotrope,
energetically near-degenerate alternate structures have been
postulated including blue phosphorus, $\gamma$-P, $\delta$-P and
their combination in one
layer \cite{{Tomanek1},{Tomanek2},{Tiling},{DT235},{polydiversity0},%
{polydiversity1},{polydiversity2},{polydiversity3},{NLribbons}}.
Identifying and discriminating between these structures requires a
tool capable of probing structures in real space with
sub-nanometer resolution. Scanning Tunneling Microscopy (STM) is
such an ideal tool, but scanning requires having chemically stable samples.

Even though phosphorene is known to degrade under ambient
conditions \cite{{ambient},{Koenig},{Hersam}}, significant progress
has been achieved in its stabilization by passivating the exposed
surface \cite{DT234}. In the initial transport studies this
passivation has been achieved by a thick PMMA
coating \cite{Koenig,Peide,Yuanbo,Buscema} that, of course,
precludes an STM study of the underlying structure. Besides PMMA,
however, monolayers of graphene and hexagonal boron nitride
({\em h}-BN)%
 \cite{Graphene_protector1,Graphene_protector2,protective1} have
proven effective in this sense, constituting the thinnest ``galvanizing agents.'' Capping by {\em h}-BN
or graphene monolayers has also been shown to protect phosphorene from
degradation \cite{Hersam,Lau}. Similar to graphene, {\em h}-BN is a
chemically inert material that can be exfoliated from bulk
materials by mechanical transfer
 \cite{hBN-mech1,hBN-mech2,hBN-mech3,hBN_subst1} or can be
synthesized by chemical vapor deposition \cite{hBN-CVD}. Unlike
in semimetallic graphene, the observed fundamental band gap of {\em h}-BN is very large and close
to 6~eV \cite{hBN_expgap}.

\begin{figure*}[tb]
\includegraphics[width=1.0\textwidth]{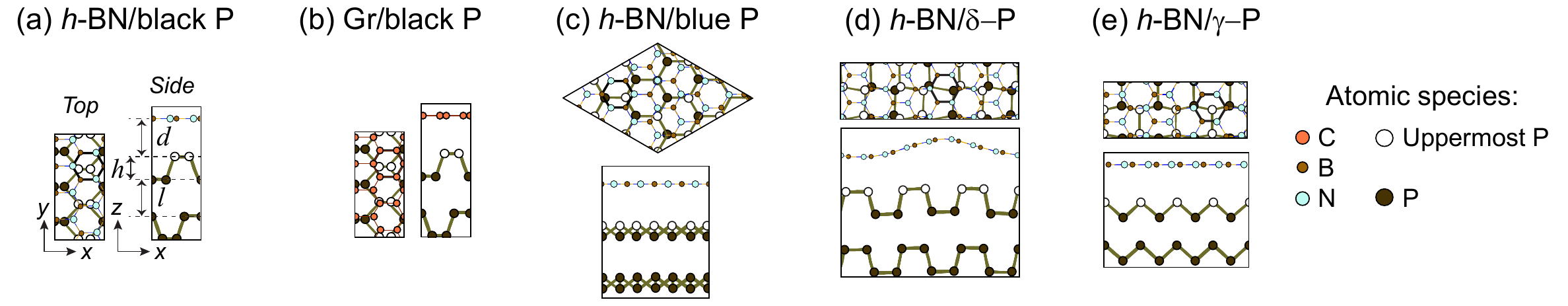}
\caption{(Color online.) Atomistic structure of
five-layer phosphorene slabs capped with hexagonal boron nitride
({\em h}-BN) or graphene (Gr) monolayers (only the two uppermost phosphorene layers and one capping monolayers are shown). As it will turn out, we will be unable to register significant phosphorene features through graphene at an equilibrium relative separation in our STM simulations. In addition, the structures capped by {\em h-}BN or graphene are quite similar (c.f. structures (a) and (b)); only their relative separation $d$ changes (c.f. Table II). For these two reasons, only one structure capped by graphene is shown on this Figure (structure (b)).}
\label{Figure1}
\end{figure*}

Even though a structure may be capped by a passivating graphene
monolayer, it has been established that the underlying structure
can be revealed by STM imaging \cite{NLus}, and the main objective of this computational
study is to establish whether different phosphorene allotropes can
also be distinguished in STM studies when covered by a monolayer
of {\em h}-BN or graphene. If so, then STM may also be able to
reveal line defects including disclinations \cite{JakobsonDefects}
and grain boundaries separating different phases in the underlying
phosphorene layer \cite{p2010}. The nature and concentration of
defects is critical for the stability of phosphorene since defects
act as nucleation sites in the material degradation process.

Calculations have been carried out using the SIESTA {\em ab
initio} simulation package \cite{SIESTA} with a cut-off energy of
300~Ry. The optB88-vdW local exchange functional \cite{KBM} was
 used to describe van der Waals interactions among successive layers. Norm-conserving
Troullier-Martins pseudopotentials \cite{Troullier} and a
double-$\zeta$ basis including polarization orbitals were employed
in calculations. Structural relaxations were performed using the
conjugate gradient method until the atomic forces were below
0.04~eV/{\AA}. The electron densities were integrated over dense
k-point meshes using an in-house program, and visualized with the
{\em Denchar} program.

\section{Results and discussion}
\subsection{Equilibrium geometry and structural stability of capped phosphorus slabs}
We first determined the
equilibrium geometry of thin slabs
of multi-layer phosphorene capped on both sides with passivating
 hexagonal boron nitride or graphene monolayers as shown in
Fig.~\ref{Figure1}. We considered five-layer
slabs of black phosphorus (
black P), blue phosphorus (
blue P), $\delta$ phosphorus ($\delta$-P) and
$\gamma$ phosphorus ($\gamma$-P)%
 \cite{Tomanek1,Tomanek2,vdW1,vdW2,vdW3} with no
relative in-plane displacements nor rotational faults. The
fundamental band gap, and the optimum inter-layer separation $l$
defined in Fig.~\ref{Figure1}, approach their limiting bulk
values at this thickness already \cite{Tomanek2}. The idea here is to create systems small enough that, at the same time, 
could be representative of bulk samples. This 
choice will thus aid experimental searches of different phases on thick slabs.

Keeping the optimum unit cell size of the 2D phosphorene slab, we
capped the slabs by a {\em h}-BN or a graphene monolayer on both
sides to suppress spurious dipole effects. The equilibrium
geometry of most structures considered is reproduced in
Fig.~\ref{Figure1}, with only two layers of the five-layer phosphorene
slabs shown in the structural side views. The coloring of
atomic species is as follows: The atoms in the
top P layer that are closest to the capping monolayers are shown in white (closest P); all
other P atoms are shown in brown. N atoms can be seen in light blue, B atoms in orange, and C atoms in
red. Black phosphorene, $\delta$-P and $\gamma$-P
have rectangular unit cells and Brillouin zones \cite{Tomanek2},
whereas blue phosphorene has a hexagonal unit cell and Brillouin
zone reminiscent of graphene \cite{Tomanek1}.

\begin{table}[tb]
\caption{Cell parameters and number of atoms per unit cell in
phosphorus allotropes capped by {\em h}-BN or graphene (Gr) monolayers.
The lattice constants are $a_g=2.49$~{\AA} for Gr and
$a_{BN}=2.52$~{\AA} for {\em h}-BN. No STM imaging of structures $\delta-$P and $\gamma-$P capped by graphene was pursued on this work.}%
\centering
\begin{tabular}{l|ccccccc}
\hline \hline
System:         & \multicolumn{2}{c}{black P}
                & {blue P}
                & \multicolumn{2}{c}{$\delta$-P}
                & \multicolumn{2}{c}{$\gamma$-P} \\
\cline{2-8} %
Cell parameters~(\AA):     & 4.58,          & 10.05 &
               {9.97}    & 16.51,         &
               5.43          & 13.60,         & 5.35         \\
{\em h-}BN mismatch~(\%):     &  1.1,          & 0.3 &
               {4.7}    & 7.0,         &
               $-$5.5          & 3.8,         & 5.9         \\
Gr mismatch~(\%):     &  5.7        & 0.9 &
               0.1    & --        &
              --          & --       & --         \\

Supercell size:   & \multicolumn{2}{l}{1$\times$3} &
                   {3$\times$3} &
                   \multicolumn{2}{l}{3$\times$1} &
                   \multicolumn{2}{l}{4$\times$1} \\
Number of atoms: & \multicolumn{2}{l}{92}  &
                        {154} &
                        \multicolumn{2}{l}{184} &
                        \multicolumn{2}{l}{128} \\
Number of P atoms:     & \multicolumn{2}{l}{60} &
                        {90} &
                        \multicolumn{2}{l}{120}&
                        \multicolumn{2}{l}{80} \\
\hline %
\end{tabular}
\label{Table1}
\end{table}

\begin{table*}[tb]
\caption{Structural information and adhesion energy $E_{ad}$ of
five-layer phosphorene slabs capped by {\em h}-BN or graphene (Gr). As
defined in Fig.~\ref{Figure1}, the average closest distance among
P atoms (in white in the structural models) and the
{\em h}-BN or graphene monolayer is $d$, the average thickness of
the P layer is $h$, and the average interlayer separation on the P
slabs is $l$. $E_{ad}/A$ is the adsorption energy $E_{ad}$ of the
capping layer on the phosphorus substrate defined in
Equation~\ref{eqEad}, divided by the area of the unit cell $A$. No $\delta-$ nor $\gamma-$structures capped with graphene were pursued on this study.}
\centering
\begin{tabular}{l|cccccc}
\hline \hline
               &{\em h}-BN/black P & Gr/black P &
               {\em h}-BN/blue P   & Gr/blue P   & {\em h}-BN/$\delta$-P &
               {\em h}-BN/$\gamma$-P \\
\cline{2-7}%
$d$ (\AA)             & 3.68  & 3.56& 3.75 & 3.60 & 3.89  & 3.55  \\
$h$ (\AA)             & 2.19  & 2.19& 1.27 & 1.27 & 2.24  & 1.61  \\
$l$ (\AA)             & 3.58  & 3.42& 3.68 & 3.58 & 3.33  & 2.40  \\
$E_{ad}/A$~(meV/{\AA}$^2$) &$-$21.60&$-$20.38&$-$20.26&$-$18.02&$-$19.49&$-$20.62\\%
\hline
\end{tabular}
\label{Table2}%
\end{table*}


The size of the supercells, listed in Table~\ref{Table1}, ranges
from 92 to 184 atoms. The average distance $d$ between the topmost
P atoms (shown in white in Fig.~\ref{Figure1}) and the {\em
h}-BN or graphene capping layer is listed in Table~\ref{Table2}
and it ranges from 3.55~{\AA} in $\gamma$-P to 3.89~{\AA} in
$\delta$-P. Since the equilibrium lattice constant of {\em h}-BN
and graphene monolayers differs in between 0.1 and 7.0\% from that of the
phosphorus slabs (Table I), the capping layers have been stretched or
compressed to enforce epitaxy. In the $\delta$-P slab, the {\em
h}-BN overlayer shows an energetic preference for a wavy
structure \cite{DT220} --shown in Fig.~\ref{Figure1}(d)-- over a
planar compressed structure, displaying a pattern reminiscent of
the observed structure in slabs of phosphorus IV \cite{PRLpiv}.

 It is important to note that not all phosphorene slabs were capped with graphene in the present study. This is so mainly because
 graphene's large non-zero electron density at relevant scanning energies precludes the use of these slabs for semiconductor
 applications, and because graphene's non-zero electronic density may obscure a simple identification of phosphorene phases underneath.

We investigated the adhesion of the capping layer on the
phosphorus slab. We defined the adhesion energy $E_{ad}$ by:
\begin{eqnarray}
E_{ad} = (1/2) [E_{tot}({\rm capped~P~slab}) - \nonumber\\
2E_{tot}({\rm capping~layer}) -  E_{tot}({\rm P~slab})]\;.
\label{eqEad}
\end{eqnarray}
Here, $E_{tot}({\rm capped~P~slab})$ is the total energy of the P
slab capped by either $h$-BN or graphene at both exposed surfaces,
$E_{tot}({\rm P~slab})$ is the total energy of the isolated P
slab, and $E_{tot}({\rm capping~layer})$ is the energy of a single
monolayer of $h$-BN or graphene. Negative values indicate energy
gain upon adhesion, and the factor $1/2$ takes care of the fact that the slab is
capped at both sides. We find it useful to divide the adsorption
energy by the interface area $A$ and list the values of $E_{ad}/A$
in Table~\ref{Table2}. {\em h}-BN/black P turns out to be the most
stable capped system, with an energy gain upon adhesion of
$21.60$~meV/{\AA}$^2$. {\em h}-BN/$\delta$-P has the weakest
adhesion energy ($19.49$~meV/{\AA}$^2$)  among {\em h}-BN capped systems, which is caused by a
larger average separation between {\em h}-BN and the wavy
phosphorene layer, as seen in Fig.~\ref{Figure1}(d). When phosphorene slabs are capped by graphene monolayers the adhesion energy becomes weaker than 
on their {\em h}-BN-capped counterparts, as evidenced by a comparison of values of $E_{ad}/A$ among pairs of columns 
(e.g., {\it h-}BN/black P and Gr/black P, or {\it h-}BN/blue P and Gr/blue P in Table \ref{Table2}). We discuss the 
electronic properties of these slabs next.

\subsection{Electronic properties of capped phosphorus slabs}

The Fermi level of capped phosphorene slabs is set at
$E_F=0$~eV. We provide band structures, the total density of states, 
and the electronic density projected onto atoms belonging to the capping monolayers in Fig.~2. 
 Graphene has a zero band-gap and the DFT value of the fundamental band gap of an isolated {\em h}-BN monolayer is 4.5~eV. We
 determine a band gap of 0.44~eV for black P, 0.84~eV for blue P, and 0.21~eV for $\delta$-P slabs, with the
$\gamma$-P slab being metallic \cite{Tomanek2} (c.f., Fig.~2). These gaps are further emphasized in the
total density of states (DOS) plots shown to the right of each band structure plot. In slabs made with two semiconductors
(Figs.~2(a), ~2(c) and ~2(e)) we set the Fermi energy at the mid-gap among phosphorene states.

The relative 
alignment of the nominal band-gap edges of {\it h-}BN is shown by thick (red) horizontal lines, and the relative energy of these band edges is clearly sensitive
to the relative separation $\Delta z$ among {\it h-}BN and the phosphorene slab underneath, as indicated by the dash horizontal line in Fig.~2(a) obtained
when {\it h-}BN pushed by 0.4 \AA{} down into black P. Additionally, an evident orbital hybridization as a function of $\Delta z$ into electronic states within 
the nominal band gap has been highlighted by the tilted arrows in Fig.~2(a). We will discuss hybridization in full detail later on.

 The gray (dashed) rectangles in all plots indicate the energy window in which STM imaging will take place; the red horizontal lines on {\it h-}BN capped 
 phosphorene allotropes indicate that the STM scanning window is located within the nominal gap of an isolated  {\it h-}BN monolayer (subplots 2(a), 
 2(c), 2(e) and 2(f)). As it is known, graphene has a non-zero electronic density in all scanning windows (subplots 2(b),
 and 2(d)).

 We indicate the total density of states and its projection onto the $p_z$ channel for isolated {\em h}-BN \cite{jpcm} and graphene in 
Figs.~2(g) and 2(h) as well, as the distribution of $p_z$ orbitals among B, N and C atoms  may shed light on the complex 
hybridization taking place when these monolayers are in close proximity to phosphorene allotropes at this energy window. While graphene 
is known to have a $\pi$ ($p_z$) orbital character around the Fermi energy (c.f.,~Fig.~2(g)), the valence (conduction) band of {\it h-}BN has a
 predominant N (B) $p_z$ character (c.f., Fig.~2(h)). 
 There is a subtle hybridization of the {\em h}-BN monolayer on these phosphorene slabs that is discussed next.

\begin{figure*}[tb]
\includegraphics[width=1.0\textwidth]{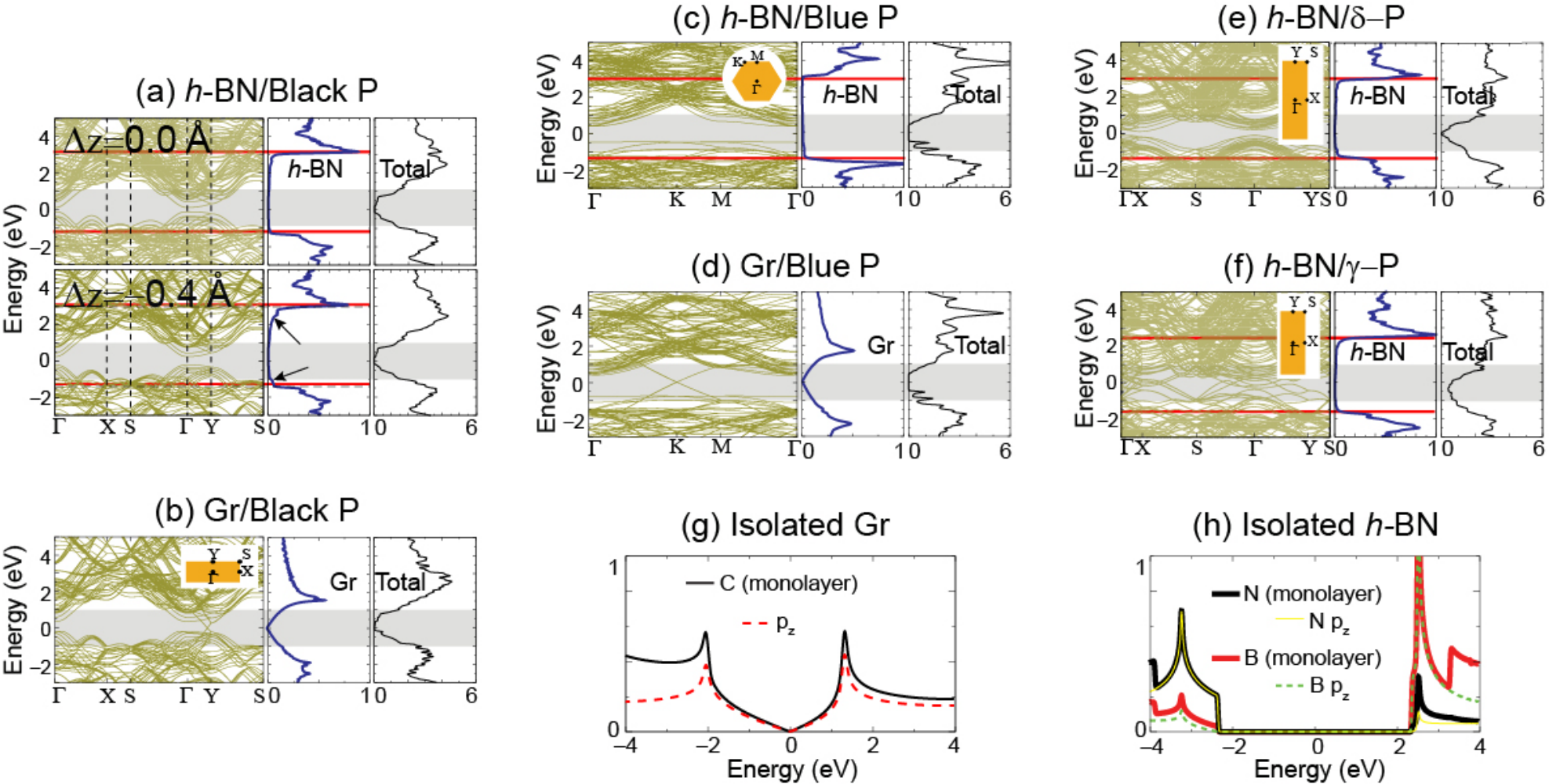}
\caption{(Color online.) Electronic structure properties of five-layer phosphorene slabs capped with hexagonal boron nitride 
({\em h}-BN, subplots a,c,e and f) and graphene (subplots b and d) monolayers. Electronic bands are in the left subpanels, and 
Brillouin zones are seen as insets of these plots. The arrows in subplot a for $\Delta z=-0.4$ \AA{} indicate 
a clear hybridization within the nominal gap of {\em h-}BN. This hybridization produces a height-dependent 
shift of the band edges that is clearly seen in subplot (a) by comparing the bold red and the thin horizontal dashed lines.
Although less visible, such hybridization occurs for all {\it h-}BN capped slabs. The 
projected {\em h}-BN and total density of states are shown in the 
right subpanels. The density of states for isolated Graphene and for {\em h}-BN, and their projections into p$_z$ orbitals are displayed in the (g) and (h) subplots.}
\label{Figure2}
\end{figure*}

\begin{figure*}[tb]
\includegraphics[width=1.0\textwidth]{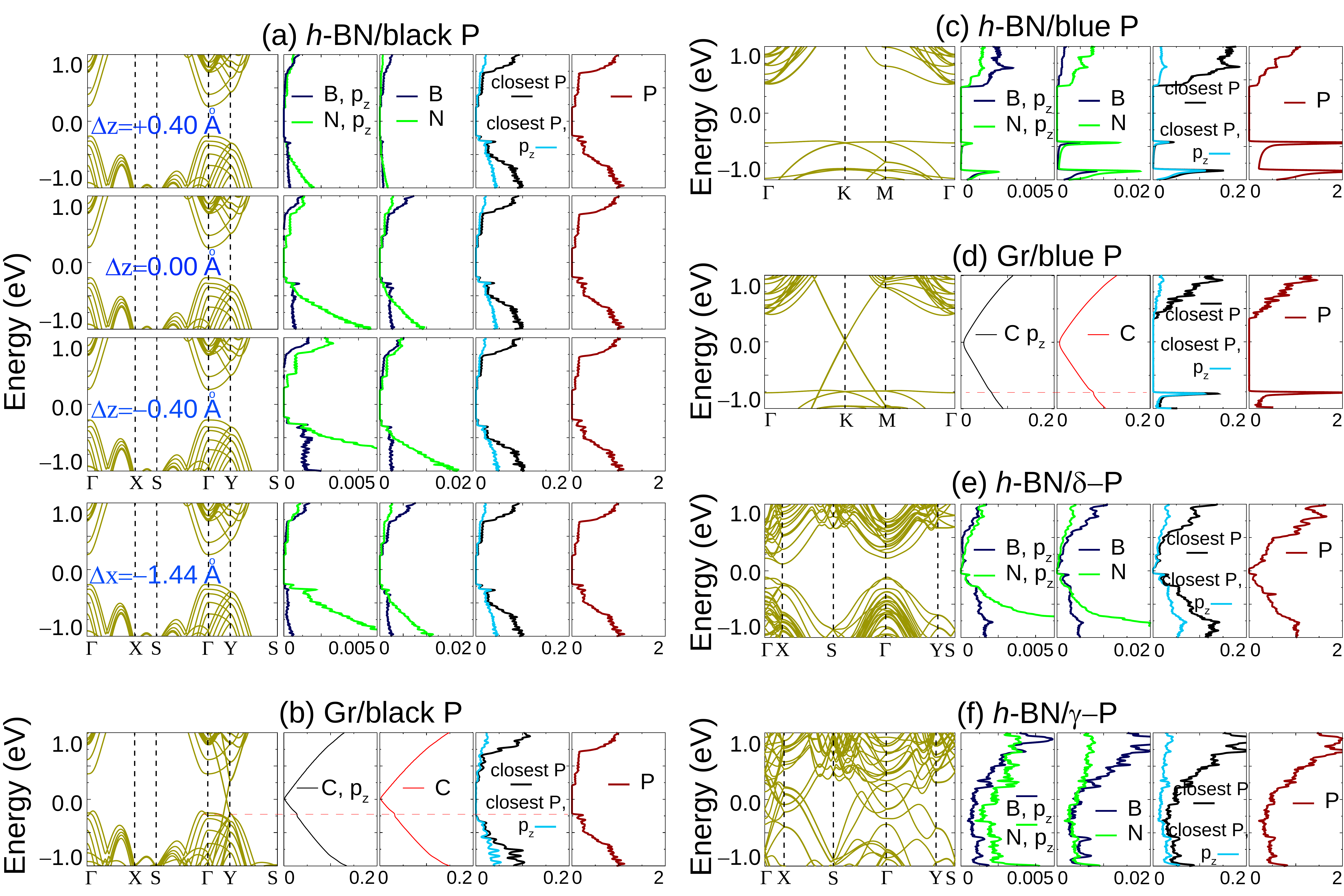}
\caption{(Color online.) Band structures and projected density of states onto B, N, C and P atoms within $\pm$ 1 eV for phosphorene slabs 
capped by {\em h}-BN and graphene monolayers. Note that {\em h}-BN hybridizes with phosphorus states within its nominal 
band gap (c.f. Table \ref{Table3}). Graphene states in capped structures exhibit an intrinsic charge that is one order of 
magnitude larger than the charge induced on {\em h}-BN by hybridization, and only shallow additional hybridization by phosphorene
is visible in subplots (b) and (d), as highlighted by horizontal dashed lines. The hybridized states shown here 
are those observed on STM images in Figs.~\ref{Figure4} to \ref{Figure7}.}
\label{Figure3}
\end{figure*}

The energy range in Fig.~2 was made sufficiently large
 in order to capture the nominal band edges of {\it h-}BN, but we have obtained STM images within $\pm$ 1 eV from the Fermi 
 energy (dashed rectangles in Fig.~2), making it important to look at the electronic states at this energy window to better understand 
 the electronic states seen on STM images. In  Fig.~3, we correlate band structure features with atomic species and orbital 
 character through projected density of states plots for energies relevant to our STM images, that will be discussed later on.  

For this purpose, the total density of P atoms is shown on the rightmost subplots on Fig.~3, and the electronic density projected onto the P atoms closest 
to the capping monolayers (seen in 
white in Fig.~1) has been recorded too. We registered the $p_z$ character of these orbitals too, finding a relatively larger $p_z$ character 
for the P orbitals with negative energies (c.f., Figs.~3(a) to 3(f)). 

 While --by definition-- an isolated {\it h-}BN monolayer has a zero electron density for energies within its electronic gap (Fig.~2(h)), this ceases to be
 the case as soon as the slab is in proximity of other material with a smaller gap: {\it h-}BN has a non-zero electronic density due to 
 hybridization with the closest P atoms (Figs.~2(a), 2(c), 2(e), and 2(f); this is more clearly seen in Figs.~3(a), 3(c), 3(e) and 3(f)). Furthermore, it 
 is crucial to emphasize the density scale of the hybridized {\it h-}BN monolayers, which is
 an order of magnitude smaller than that set for the closest P atoms and graphene (the density scale of electron per 
 unit cell is consistent among Figs.~2 and 3). The density scale for graphene
 is identical to that of the closest P atoms, and this could preclude an easy identification of P-related features for graphene-capped phosphorene. One notes that
 the hybridization seen on {\it h-}BN follows the same pattern seen in Fig.~2(h): Namely, states with negative energies have a predominant N-character, 
 while states with positive energies display a larger B-character. The only exception to this trend is realized in blue phosphorene, subplot 3(c) which indicates
 the role of structure and symmetries on the resulting hybridization. 
 An additional piece of information concerns the orbital projection onto $p_z$ orbitals on {\it h-}BN, which is almost an order of magnitude smaller than the
 overall density recorded on {\it h-}BN at these energies, and implying contributions from other orbitals to these hybridized states as well 
 (c.f., Table \ref{Table3}).

\begin{table*}[tb]
\caption{Total charge $\Delta Q$ in electrons per unit cell, and percentual charge distribution into $s$ and $p$ orbitals within a $-1$ to 1 V bias window for phosphorene slabs capped by hexagonal boron nitride ({\it {h}}-BN) or graphene (Gr) monolayers. The hybridization on {\it {h}}-BN under forward bias is mostly localized on N p$_z$ channels. Graphene-capped materials have a large intrinsic charge on the C p$_z$ orbitals that will mask the electronic density of the P slabs underneath on STM images.}
\centering
\begin{tabular}{c|m{1.1cm}m{1.1cm}m{1.1cm}m{1.1cm}m{1.1cm}m{1.1cm}m{1.1cm}m{1.1cm}m{1.1cm}m{1.1cm}m{1.1cm}m{1.1cm}}
\hline \hline
System:  & \multicolumn{4}{c}{{\em h-}BN/black P, $\Delta z=+0.4$ \AA} & \multicolumn{4}{c}{{\em h-}BN/black P, $\Delta z=0.0$ \AA} & \multicolumn{4}{c}{{\em h-}BN/black P, $\Delta z=-0.4$ \AA} \\ \hline
Bias window (V): & \multicolumn{2}{c} {$(-1,0)$} & \multicolumn{2}{c} {$(0,+1)$} & \multicolumn{2}{c} {$(-1,0)$} & \multicolumn{2}{c} {$(0,+1)$} & \multicolumn{2}{c}{$(-1,0)$} & \multicolumn{2}{c} {$(0,+1)$} \\
$\Delta Q$ (e/uc): & \multicolumn{2}{c}{0.001} & \multicolumn{2}{c}{0.001} & \multicolumn{2}{c}{0.004} & \multicolumn{2}{c}{0.002} & \multicolumn{2}{c}{0.006} & \multicolumn{2}{c}{0.003} \\
Atomic species: & B & N ~&~ B & N & B & N ~&~ B & N & B & N ~&~ B & N \\
$s$ (\%): & 16 & 3 ~&~ 16 & 1  & 10 & 6 ~&~ 3 & 2 & 9 & 6 ~&~ 1 & 2 \\
$p_x$ and $p_y$ (\%): & 20 & 0 ~&~ 0 & 0 & 15 & 0 ~&~ 0 & 1 & 12 & 1 ~&~ 0 & 3 \\
$p_z$ (\%): & 14 & 17 ~&~ 20 & 52 & 12 & 31 ~&~ 15 & 60 & 13 & 35 ~&~ 15 & 62 \\ \hline \hline
System: & \multicolumn{4}{c}{{\em h-}BN/black P, $\Delta x=+1.44$ \AA} & \multicolumn{4}{c}{Gr/black P} & \multicolumn{4}{c}{{\em h-}BN/blue P} \\ \hline
Bias window (V): & \multicolumn{2}{c}{$(-1,0)$} & \multicolumn{2}{c}{$(0,+1)$} & \multicolumn{2}{c}{$(-1,0)$} & \multicolumn{2}{c}{$(0,+1)$} & \multicolumn{2}{c}{$(-1,0)$} & \multicolumn{2}{c}{$(0,+1)$} \\
$\Delta Q$ (e/uc): & \multicolumn{2}{c}{0.003} & \multicolumn{2}{c}{0.003} & \multicolumn{2}{c}{0.092} & \multicolumn{2}{c}{0.105} & \multicolumn{2}{c}{0.001} & \multicolumn{2}{c}{0.003} \\
Atomic species: & B & N ~&~ B & N & \multicolumn{2}{c}{C} & \multicolumn{2}{c}{C} & B & N ~&~ B & N \\
$s$ (\%): &  12 & 5 ~&~ 6 & 2 & \multicolumn{2}{c}{0} & \multicolumn{2}{c}{0} & 7 & 0 ~&~ 9 & 3 \\
$p_x$ and $p_y$ (\%): & 17 & 1 ~&~ 1 & 1 & \multicolumn{2}{c}{0} & \multicolumn{2}{c}{0} & 26 & 0 ~&~ 20 & 0 \\
$p_z$ (\%): & 19 & 17 ~&~ 6 & 78 & \multicolumn{2}{c}{89} & \multicolumn{2}{c}{86} & 30 & 15 ~&~ 18 & 20 \\ \hline \hline
System: & \multicolumn{4}{c}{Gr/blue P} & \multicolumn{4}{c}{{\em h-}BN/Delta P} & \multicolumn{4}{c}{{\em h-}BN/Gamma P} \\ \hline
Bias window (V): & \multicolumn{2}{c}{$(-1,0)$} & \multicolumn{2}{c}{$(0,+1)$} & \multicolumn{2}{c}{$(-1,0)$} & \multicolumn{2}{c}{$(0,+1)$} & \multicolumn{2}{c}{$(-1,0)$} & \multicolumn{2}{c}{$(0,+1)$} \\
$\Delta Q$ (e/uc): & \multicolumn{2}{c}{0.076} & \multicolumn{2}{c}{0.083} & \multicolumn{2}{c}{0.013} & \multicolumn{2}{c}{0.003} & \multicolumn{2}{c}{0.009} & \multicolumn{2}{c}{0.021} \\
Atomic species: & \multicolumn{2}{c}{C} & \multicolumn{2}{c}{C} & B & N ~&~ B & N & B & N ~&~ B & N \\
$s$ (\%): & \multicolumn{2}{c}{0} & \multicolumn{2}{c}{0} & 11 & 5 ~&~ 0 & 0 & 4 & 0 ~&~ 3 & 0 \\
$p_x$ and $p_y$ (\%): & \multicolumn{2}{c}{0} & \multicolumn{2}{c}{0} & 14 & 3 ~&~ 1 & 3 & 9 & 0 ~&~ 7 & 0 \\
$p_z$ (\%): & \multicolumn{2}{c}{86} & \multicolumn{2}{c}{86} & 13 & 23 ~&~ 10 & 69 & 12 & 14 ~&~ 10 & 22\\ \hline
\end{tabular}
\label{Table3}
\end{table*}

Electronic hybridization implies charge transfer: A slight electron/hole charge transfer sets the Fermi energy right at the conduction/valence 
band edge in an isolated system. But 
$h-$BN is not isolated in Fig.~1; it is in close proximity to phosphorene slabs that induce a small electronic hybridization 
within its otherwise empty nominal electronic gap. This slight hybridization of electronic states within 
the nominal gap of {\em h}-BN is documented on subplots B and N in Figs.~3(a), 3(c), 3(e) and 3(f), as well as in Table \ref{Table3}: Phosphorene slabs transfer a
small electronic charge $\Delta Q$ to the {\it h-}BN monolayer, and this 
orbital hybridization allows for a gradual band shift of {\em h}-BN with respect to the phosphorene slab as a function of $d$. This
effect will be explored further in the next paragraph, when $h-$BN is intentionally pulled towards/away from the black phosphorene slab, but  a
similar slight hybridization was reported in weakly adsorbed carbon nanotubes on Si some time ago \cite{JAP2006}.

 When an STM samples a layered material, it pushes/pulls it away
from its equilibrium atomistic configuration as it attempts to
establish a feedback current \cite{NLus}, so
the alignment of the {\em h}-BN electron bands with respect to
those of the phosphorene slabs  further depends on the relative
distance $d$ between {\em h}-BN and the phosphorene slab.
 This effect is studied here in detail: We register small band shifts of the {\em h}-BN
with magnitudes $-0.265$~eV, $-0.166$~eV, $-0.075$~eV, $+0.045$~eV, and
$+0.075$~eV with respect to the band alignment shown in
Fig.~2a, upper subplot ($\Delta z=0.0$ \AA), as the {\em h}-BN monolayer is vertically displaced by a distance $\Delta z=-3.0$, 
$-0.4$, $-0.2$, $+$0.2, or $+$0.4~{\AA} away from
the black P slab, with respect to its equilibrium atomistic
separation $d$ in Fig.~1a that has a value of $\Delta z=0$ \AA{}. The dashed horizontal lines seen with the naked eye on Fig.~2(a), 
subplot $\Delta z=-0.4$ \AA{} confirm this observation. Structural defects or chemical 
contamination will cause additional relative band shifts.

Additional details on the mechanism responsible for the slight band shift seen on Fig.~2(a) can be found by contrasting subplots $\Delta z=+0.4$, 
$0.0$, and $-0.4$ \AA{} in Fig.~3(a): It is clear from these Figures that the hybridization on B and N is largest --implying the larger charge transfer
that is reported on Table \ref{Table3}-- the closest they 
are to the phosphorene slab ($\Delta z=-0.4$ AA); this is emphasized by using the same scale in the PDOS for {\em h-}BN for all subplots in Fig.~3.

In Table \ref{Table3}, we integrate the electronic charge $\Delta Q$ onto B, N or C 
electronic orbitals within 1 V bias windows from $E_F$. $\Delta Q$ is certainly non-zero, ranging in between 0.001 and 0.021 electrons per unit cell for 
{\em h-}BN capped slabs, and between 0.076 and 0.105 electrons per unit cell for graphene-capped slabs. The slight difference in charge 
for graphene arises mainly from the well-known Fermi velocity renormalization, due to the different magnitudes of strain employed 
to enforce epitaxy in black P and in blue P (c.f., Table \ref{Table1}). 
The important points are: (i) the total charge within the nominal 
gap for {\it h-}BN is non-zero due to orbital hybridization; (ii) graphene has an intrinsic charge that is an order of magnitude 
higher, which will preclude a clear-cut observation of P-related features within the STM scanning energy window (phosphorene features may be visible at the energies 
shown by horizontal dashed lines in Figs.~3(b) and 3(d); still, it will be simpler to tell phsophorene allotropes through {\it h-}BN, as it will be shown later on).

 The phosphorus atom wavefunctions hybridize the orbitals of the capping layer. For {\em h}-BN, it is precisely these hybridized 
 orbitals that become observable in STM images to be shown later on. Graphene has an electronic density that is comparable with that of 
 the closest phosphorene atoms: Since the STM image is captured over graphene and farther away from the phosphorene slabs, graphene contributes
 a larger electron density in these systems; this is at variance with what is observed on III-V semiconductors, where surface states still overcome 
 the density of graphene at the relevant scanning energies \cite{NLus}. (The effect is material-dependent, as it may be expected.)

 Additional information in Table \ref{Table3} indicates how the charge projects among the $s$ and $p$ orbitals of the capping slabs. Without exception, 
 all slabs capped with {\it h-}BN have a larger density on their N $p_z$ orbitals while under forward bias (negative energies) as it was seen in Fig.~2(h) and
 Fig.~3. In the case of reverse bias, this preponderance is not as marked: One even observes a slightly larger density onto the B $p_z$ orbital 
 when {\it h-}BN is horizontally displaced by  $\Delta x=1.44$ \AA{} on black P, as the registry among P and B atoms is enhanced there.

On graphene-capped monolayers, the $s$ and the in-plane $p$ orbitals contribute no electron density, while almost 90\% of the density is projected 
into $p_z$ orbitals that form the carbon $\pi-$bond, and we register a slight $\sim 10$\% contribution from the $d-$orbitals in the atomic-based basis 
set to such density (c.f., Table \ref{Table3}). Although the $p_z$ contribution in hybridized {\it h-}BN is still dominant, non-zero contributions 
from the $s$, $p_x$ and $p_y$ orbitals can be registered at these biases. Additionally, the contribution of $d-$ orbitals to the total charge can be 
computed by substracting 100\% to the sum of values in Table \ref{Table3}: For instance, the contribution of $d-$orbitals is 19\% (26\%) under forward 
(reverse) bias for the {\it h-}BN/Black P system with $\Delta z=0.0$ \AA. While protecting phosphorene from chemical degradation, {\it h-}BN will permit 
the identification of atomistic features
of the phosphorene structure, as will be discussed later.

\subsection{Simulation of STM images}

It is well established that STM images can be simulated by {\em ab
initio} density functional calculations \cite{DT029}. Applying a
small bias voltage $V_{bias}$ between the sample and the STM tip
yields a tunneling current, whose density $j({\bf r})$ can be
obtained from a simple extension \cite{STMSelloni85} of the
expression derived by Tersoff and
Hamann \cite{{STMPRL83},{STMPRB85}}:
\begin{equation}
j({\bf r},V_{bias}){\propto}\rho_{STM}({\bf r},V_{bias})\;,%
\end{equation}
where
\begin{equation}
\rho_{STM}({\bf r},V_{bias})=\int_{E_F-eV_{bias}}^{E_F}
                             dE\rho({\bf r},E)%
\label{eq2}
\end{equation}
and
\begin{equation}
\rho({\bf r},E)=\sum_{n,{\bf k}} |\psi_{n{\bf k}}({\bf r})|^2
                \delta(E_{n,{\bf k}}-E) \;.
\label{eq3}
\end{equation}
Here, $\rho({\bf r},E)$ is the local density of states at the
center of curvature of the tip at ${\bf r}$ and
$\psi_{n{\bf{k}}}({\bf r})$ are the electron eigenstates of the
unperturbed surface at energy $E_{n,{\bf k}}$. These eigenstates
are commonly represented by Kohn-Sham eigenstates obtained using
density functional theory. The implied
assumptions \cite{{STMSelloni85},{STMPRL83},{STMPRB85}} are that
the relevant tip states are well described by $s$ waves with a
constant density of states. Furthermore, the tunneling matrix
element is considered to be independent of the lateral tip
position for a constant tip-to-surface distance and also
independent of the bias voltage $V_{bias}$ in the narrow (but
nonzero) energy region $[E_F-eV_{bias},E_F]$. Whereas
Equation~\ref{eq2} describes tunneling from occupied states of the
sample to the tip, an obvious change of this expression can
describe tunneling from the tip to unoccupied states in the energy
range $[E_F,E_F+e|V_{bias}|]$. This latter configuration is
referred to as reverse bias and described by a negative bias
voltage.

\subsection{Simulated STM images of bare few-layer phosphorene slabs}

\begin{figure}[bt]
\includegraphics[width=0.48\textwidth]{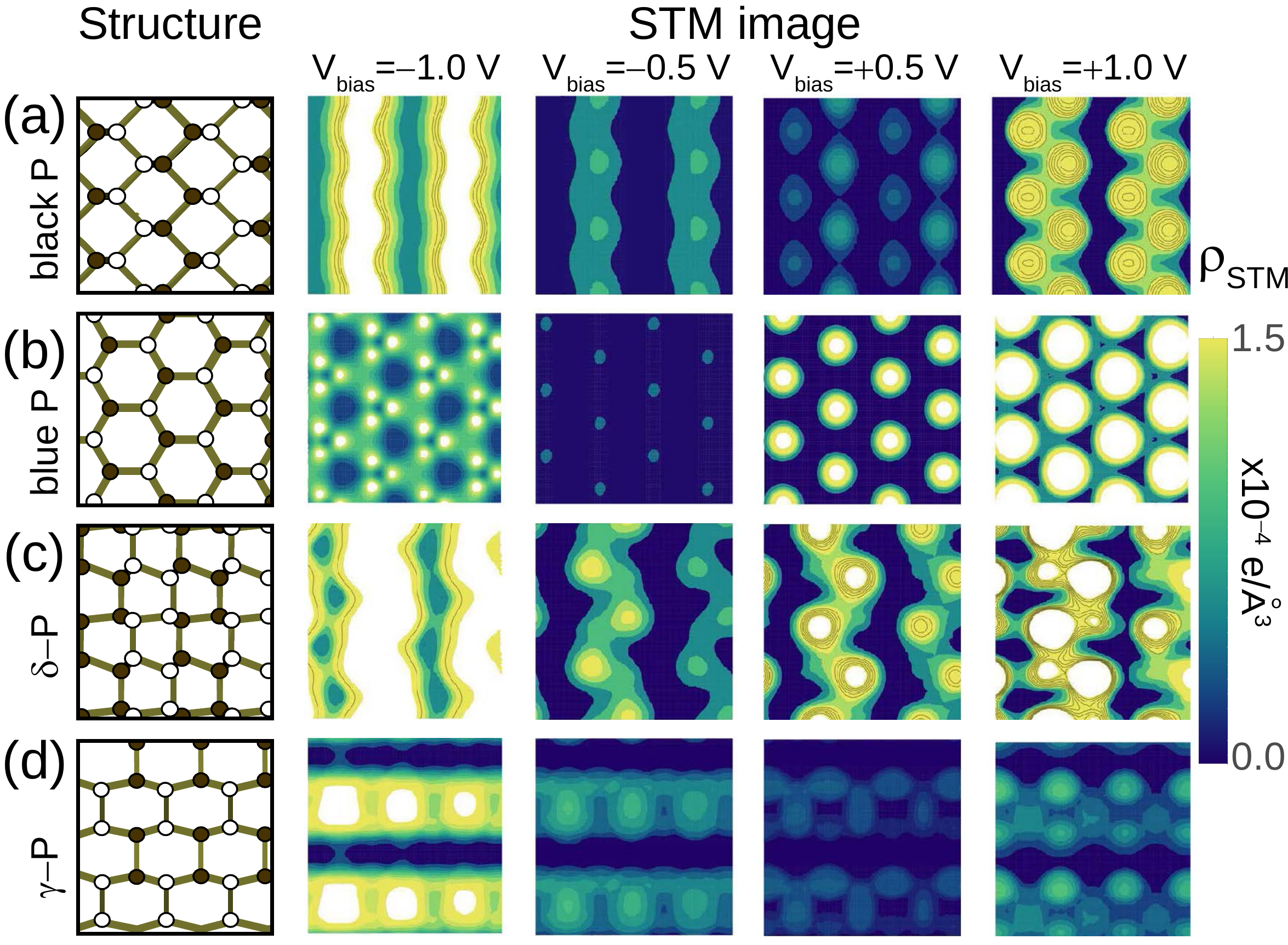}
\caption{(Color online.) Simulated STM images of a 10~{\AA}$\times$10~{\AA} area
of bare (a) black phosphorene, (b) blue-phosphorene, (c)
$\delta$-P and (d) $\gamma$-P slabs. Presented is $\rho_{STM}$ at
constant height, corresponding to the current imaging mode, at
different values of the bias voltage $V_{bias}$. For the sake of
simple comparison, the geometry of the topmost two layers of the
slabs is reproduced to scale in the left panels, with topmost
atoms shown in white. Streaks seen on black phosphorene,
$\delta$-P and $\gamma$-P reflect their ``ridged'' atomistic
structure, also highlighted in the structural models. The
different threefold symmetry of blue phosphorene with a triangular
sublattice, associated with the white-colored P atoms in the
topmost layer, is clearly reflected in the STM images in (b).}
\label{Figure4}
\end{figure}
\begin{figure}[t]
\includegraphics[width=0.48\textwidth]{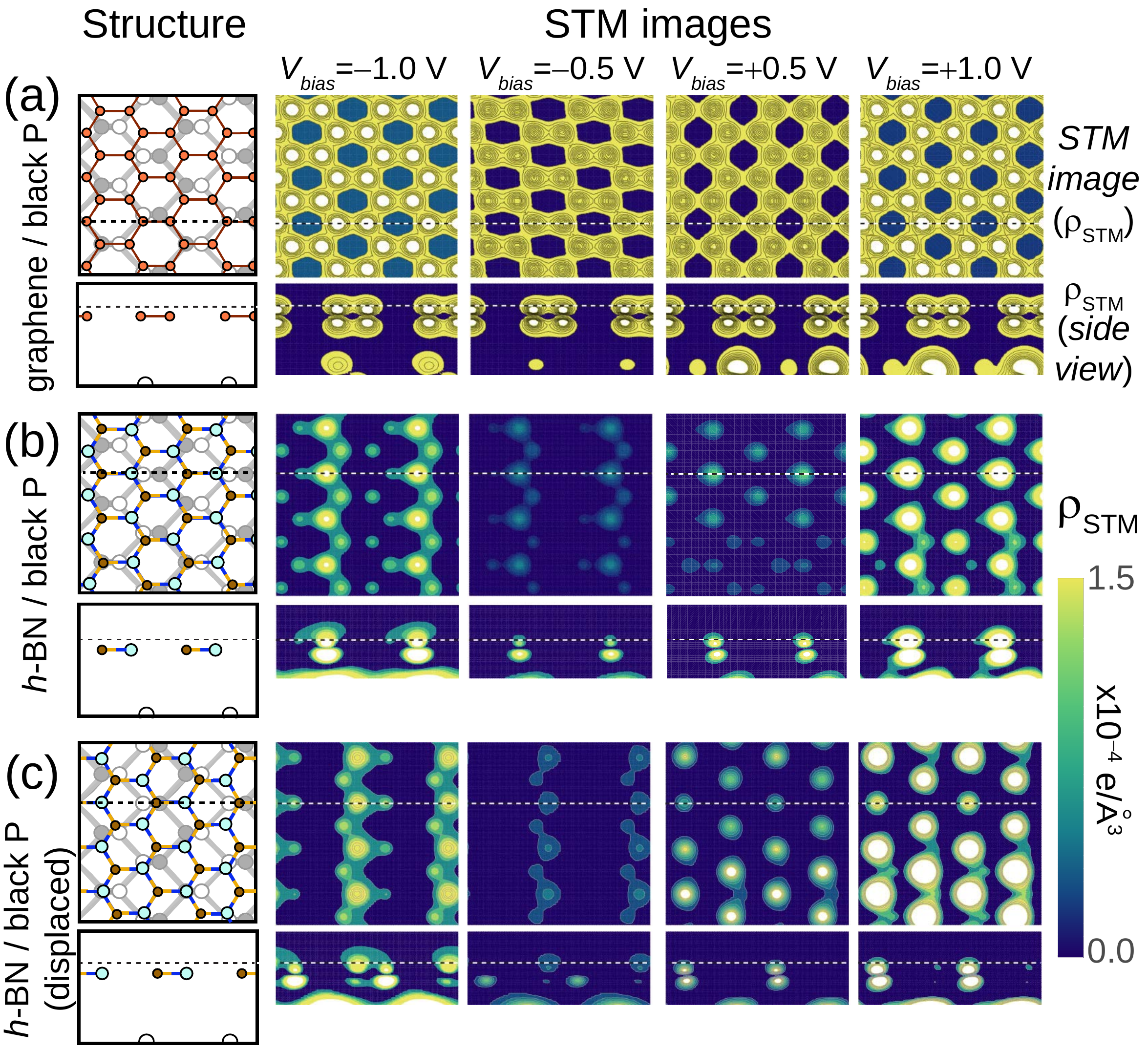}
\caption{(Color online.) Simulated STM images of a 10~{\AA}$\times$10~{\AA} area
of a black phosphorene slab capped by a monolayer of (a) graphene,
(b) {\em h}-BN and (c) a horizontally displaced {\em h}-BN
monolayer. Presented is $\rho_{STM}$ at constant height,
corresponding to the current imaging mode, at different values of
the bias voltage $V_{bias}$. The lower sub-panels represent
$\rho_{STM}$ in a plane normal to the surface. For the sake of
simple comparison, the geometry of the topmost layers of the slabs
is reproduced to scale in the left panels, with topmost P atoms
shown in white. The ridged structure of black P, which is clearly
visible in Fig.~\ref{Figure2}a at $V_{bias}=-1.0$~V, is
completely obscured by a graphene layer in (a), but is visible
underneath an {\em h}-BN layer in (b) and (c). }
\label{Figure5}
\end{figure}

\begin{figure}[t]
\includegraphics[width=0.48\textwidth]{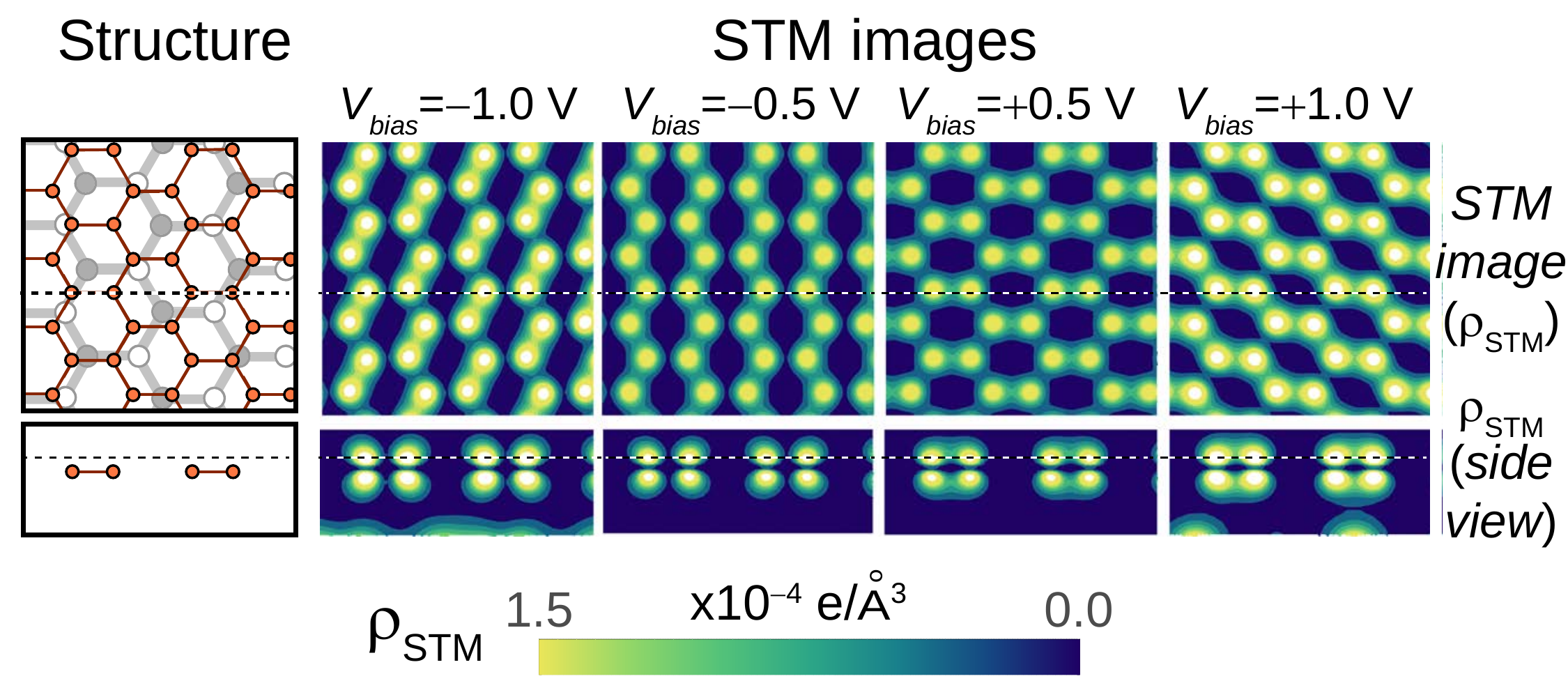}
\caption{(Color online.) Simulated STM images of a 10~{\AA}$\times$10~{\AA} area
of a blue phosphorene slab capped by a graphene monolayer. }
\label{Figure6}
\end{figure}

\begin{figure}[t]
\includegraphics[width=0.48\textwidth]{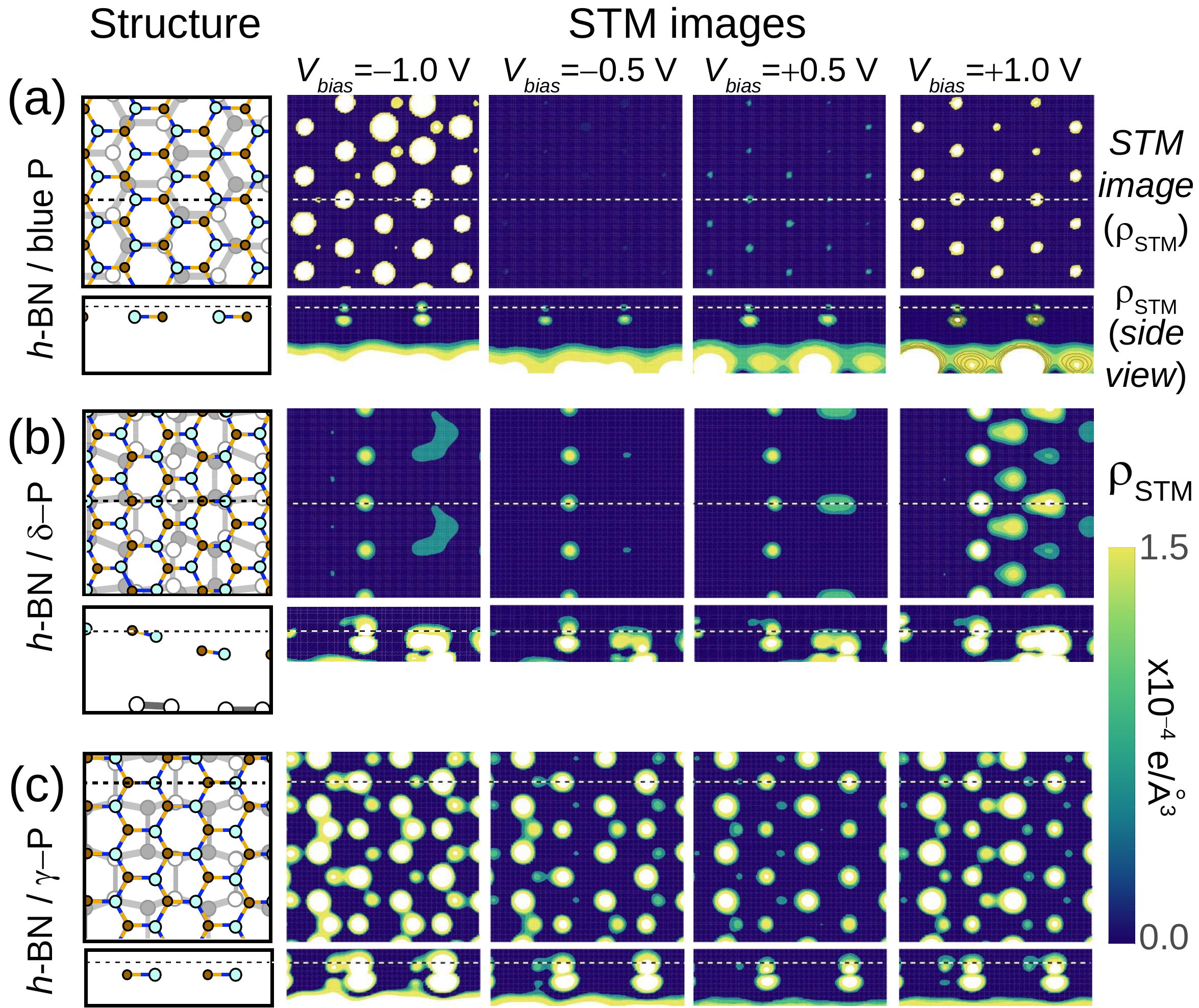}
\caption{(Color online.) Simulated STM images of a 10~{\AA}$\times$10~{\AA} area
of a (a) blue phosphorene, (b) $\delta$-P and (c) $\gamma$-P slab
capped with an {\em h}-BN monolayer. Presented is $\rho_{STM}$ at
constant height, corresponding to the current imaging mode, at
different values of the bias voltage $V_{bias}$. The lower
sub-panels represent $\rho_{STM}$ in a plane normal to the
surface. For the sake of simple comparison, the geometry of the
topmost layers of the slabs is reproduced to scale in the left
panels, with topmost P atoms shown in white.}
\label{Figure7}
\end{figure}

Simulated STM images of bare few-layer phosphorene slabs,
representing the charge density $\rho_{STM}({\bf r},V_{bias})$,
given by Equation~\ref{eq2}, at various bias voltages $V_{bias}$
are presented in Fig.~\ref{Figure4}. For better interpretation
of the images, we reproduced the atomic arrangement in the topmost
layers of the slab in the left panels. Visual comparison of the
structures and the STM images reveals that the highest values of
$\rho_{STM}$ correlate well with the location of the topmost
P atoms, which are shown in white in the structural depiction.
Underlying all STM images is a dense $k$-point mesh used in
Equation~\ref{eq3} for the charge density integration. The STM
image is represented by $\rho_{STM}$ scanned within a plane at the
distance $d=2.3$~{\AA} from the closest phosphorus atoms shown in
white. For the sake of simple comparison, we keep the scan area
and the color bars the same in Figs.~\ref{Figure4} to
\ref{Figure7}.

An antibonding-like feature is visible by the spatial distribution
localized over P atoms at forward bias $V_{bias}$ on black P
(Fig.~\ref{Figure4}(a)). On the other hand, a bonding-like feature
is reproduced under a negative bias as the electronic density is
larger at bonds between pairs of P atoms. In blue phosphorus
(Fig.~\ref{Figure4}(b)) a triangular shape of the electronic
density is seen under forward bias due to the electronic density
coming only from the upper P atoms. The images under a reverse
bias  $V_{bias}=-1 V$ show a decorated triangular lattice
displaying three bright small circles around each uppermost
(white) P atom. The higher intensity for $V_{bias}>0$ correlates
with the rather flat valence band seen on Fig.~\ref{Figure3}(c)
(the reader must divide the energy scale in Figs.~\ref{Figure2} and \ref{Figure3}
by $-|e|$ to obtain $V_{bias}$ on Figs.~\ref{Figure4} to
\ref{Figure7}).

Two vertical trenches can be seen in  $\delta$-P
(Fig.~\ref{Figure4}(c)) for forward and reverse biases, but the
shape of these trenches is different depending on bias direction:
The electronic density $\rho_{STM}$ is localized onto the upper P
atoms for positive $V_{bias}$, and it appears more distributed
over bonds under reverse biases; this is particularly evident for
$V_{bias}=-1 V$. The spatial density acquired above the $\gamma$-P
slab  also exhibits two horizontal trenches under negative bias in
a bonding-like fashion as the charge is distributed away from
atoms and into covalent bonds (Fig.~\ref{Figure4}d). Due to the
antibonding-like distribution, these electronic density trenches
can be seen localized over the P atoms when $V_{bias}>0$ $V$.

Clearly, different structural phases can be distinguished with an
STM. Next we will study if these structures remain distinguishable
by STM when covered by a capping monolayer.

\subsection{Simulated STM images of few-layer phosphorus slabs
capped by {\em h}-BN or graphene}

Simulated STM images of few-layer black phosphorene slabs capped
by {\em h}-BN or graphene are presented in Fig.~\ref{Figure5}.
In the current imaging mode, the STM current is represented by the
charge density $\rho_{STM}({\bf r},V_{bias})$, given by
Equation~\ref{eq2}. We present $\rho_{STM}$ at various bias
voltages $V_{bias}$ in a plane parallel to the surface, indicated
by a dashed horizontal line in the structural image in side view.
We also present $\rho_{STM}$ in a plane normal to the surface,
indicated by a dashed horizontal line in the structural image in
top view. These images allow us to judge, whether the structure of
a phosphorene slab may be distinguished underneath a capping
monolayer of {\em h}-BN or graphene.

The STM images in Fig.~\ref{Figure5}(b) and
\ref{Figure5}(c) indicate that the {\em h}-BN
layer allows, to some degree, to discriminate between different phosphorene
phases \cite{Tiling}. As a result of the small but finite
hybridization, the white/yellow streaks in STM images in
Fig.~\ref{Figure4}a seen at $V_{bias}=-1.0$~V --highlighting the ridges
of a black phosphorene surface-- become visible underneath
the {\em h}-BN layer in Figs.~\ref{Figure5}(b) and \ref{Figure5}(c).

In the case of III-V semiconductors, a capping monolayer of
graphene, when pushed by the STM tip to close proximity of the
substrate, was found to not to conceal the atomic structure of the
substrate \cite{NLus}. That experimental observation made us consider whether phosphorene could be seen 
through graphene/{\em h-}BN.  As seen in Fig.~\ref{Figure4}(a), this is clearly not
the case for black phosphorene covered by graphene. Independent of
the bias voltage, the simulated STM images are dominated by the
structure of the graphene overlayer. The significantly larger
density of graphene states in the scanning plane efficiently masks
the ridged structure of black phosphorene, which is most clearly
visible in the STM image of the bare surface at $V_{bias}=-1.0$~V
on Fig.~\ref{Figure5}(a). Also, the side view images -- cut along
a carbon covalent bond -- show a symmetric distribution of the
electronic density of the $\pi$-bond among pairs of carbon atoms.
 In addition, one sees in Fig.~\ref{Figure6} an overwhelming contribution from graphene states as well.
We could push graphene to the slab and hope to see a similar effect to that reported in Ref.~\cite{NLus}, 
but there may be a simpler way to protect phosphorene and determining the layered phase at the exposed surface: In 
contrast to the semimetallic graphene, {\em h}-BN will indeed permit the identification of the phosphorene allotrope 
underneath. Based on these results, we did not consider graphene-capped $\delta-$P and $\gamma-$P here.

To further understand the differences between
Figs.~\ref{Figure4}(a) and \ref{Figure5}(b), caused by the presence
of the capping {\em h}-BN layer, we invite the reader to observe the contributions from hybridized B and N 
atoms onto the STM images, as given by Fig.~\ref{Figure3} and Table \ref{Table3}.

Even though the adhesion of the {\em h}-BN layer on the
phosphorene substrate is substantial according to
Table~\ref{Table2}, contact with the STM tip may shift the capping
layer horizontally. In the following, we discuss the effect of
such a horizontal displacement on the STM image.
Figs.~\ref{Figure5}(b) and \ref{Figure5}(c) both represent a black P
slab covered by {\em h}-BN; the capping layer in \ref{Figure5}(c)
has been shifted from its position in \ref{Figure5}(b). This
displacement can be visualized most easily in the side view of the
atomic structure, in a plane normal to the surface. As seen at the
top of the left panels in Figures~\ref{Figure5}(b) and
\ref{Figure5}(c), this plane contains two P atoms in the topmost P
layer. At the bottom of the left panels of these sub-figures, we
can see the effect of a horizontal displacement by 1.44~{\AA}. One
of the P atoms, which was underneath an N atom in
Fig.~\ref{Figure5}(b), appears underneath a B atom in
\ref{Figure5}c. Even though this displacement represents the
worst-case scenario of switching from P-N registry to P-B
registry, the STM images in Figs.~\ref{Figure5}(b) and
\ref{Figure5}c are rather similar at all bias voltages, indicating
image stability with respect to displacements of the capping
layer. In particular, the ridged structure, which dominated the
image of the bare surface in Fig.~\ref{Figure4}(a), still appears pronounced
on Figs.~\ref{Figure5}(b) and
\ref{Figure5}(c).

 It is time to establish from the STM images, whether graphene allows the identification of phosphorene allotropes underneath. One can only see graphene-like features in 
 Figs.~5(a) and Fig.~6. This means that, even if graphene hybridizes with phosphorene states (dashed horizontal lines on Figs.~3(b) and 3(d)), telling blue P 
 from black P does not seem straightforward when these slabs are capped by a graphene monolayer. On the other hand,
  {\it h-}BN will permit identification of phosphorene allotropes, as it will be shown next.

We simulated STM images
of blue P, $\delta$-P and $\gamma$-P to judge whether the different allotropes can be distinguished
underneath an {\em h}-BN monolayer. We find these images,
presented in Fig.~\ref{Figure7}, to be all different, and to
also differ from those of black P in Fig.~\ref{Figure5}(b) and
\ref{Figure5}(c). In comparison to black P with a rectangular unit
cell, Fig.~\ref{Figure7}(a) clearly reveals the honeycomb lattice
of the underlying blue phosphorene.
Similarly, $\rho_{STM}$ of a capped $\delta$-P slab in
Fig.~\ref{Figure7}(b) displays two vertical stripes at locations
above N atoms at negative bias values, in close agreement with
Fig.~\ref{Figure4}(c). Due to its wavy morphology, the capping
layer binds least to the substrate and is most likely to be
displaced by an STM tip during scanning. The STM images of a
capped $\gamma$-P slab, shown in Fig.~\ref{Figure7}(c), display
similar ridges as seen in the bare $\gamma$-P slab, shown in
Fig.~\ref{Figure4}(d).

Further effects, such as relative rotations and moir{\'e} patterns, and removing the in-plane 
strain we imposed to have sufficiently small supercells for computations, will compose actual experimental images.
Nevertheless, the main message of our simulated STM images, presented in
Figs.~\ref{Figure4}-\ref{Figure7}, is that STM appears capable of
distinguishing between different phases of layered phosphorus
underneath a passivating hexagonal boron nitride monolayer and it thus represents 
the only route known at this moment to tell these phases from one another with a local probe.

\section{Conclusions}

In summary, we have studied the electronic structure and simulated
scanning tunnel microscopy of few-layer phosphorus allotropes
capped by {\em h}-BN and graphene monolayers using density
functional theory with vdW corrections. Our results indicate that
capping by {\em h}-BN does permit identifying phosphorus phases
underneath. Due to the vanishing band gap of graphene, the charge
density of capping graphene monolayers masks the structure
underneath, making this structure unsuitable to discriminate
between different phosphorene allotropes. These results may assist
the nascent experimental searches for different structural phases
of layered phosphorus.

We thank NSF-XSEDE (Grant TG-PHY090002; {TACC's \em Stampede}) and
Arkansas ({\em Razor II}) for computational support. P.R. and
S.B.L. acknowledge funding from the Arkansas Biosciences
Institute. Z.Z., J.G. and D.T. acknowledge support by the National
Science Foundation Cooperative Agreement No. EEC-0832785, titled
``NSEC: Center for High-Rate Nanomanufacturing''.


\begin{thebibliography}{61}
\expandafter\ifx\csname natexlab\endcsname\relax\def\natexlab#1{#1}\fi
\expandafter\ifx\csname bibnamefont\endcsname\relax
  \def\bibnamefont#1{#1}\fi
\expandafter\ifx\csname bibfnamefont\endcsname\relax
  \def\bibfnamefont#1{#1}\fi
\expandafter\ifx\csname citenamefont\endcsname\relax
  \def \citenamefont#1{#1}\fi
\expandafter\ifx\csname url\endcsname\relax
  \def\url#1{\texttt{#1}}\fi
\expandafter\ifx\csname urlprefix\endcsname\relax\def\urlprefix{URL }\fi
\providecommand{\bibinfo}[2]{#2}
\providecommand{\eprint}[2][]{\url{#2}}

\bibitem[{ \citenamefont{Novoselov et~al.}(2005) \citenamefont{Novoselov, Jiang,
  Schedin, Booth, Khotkevich, Morozov, and Geim}}]{Novoselov2D}
\bibinfo{author}{\bibfnamefont{K.~S.} \bibnamefont{Novoselov}},
  \bibinfo{author}{\bibfnamefont{D.}~\bibnamefont{Jiang}},
  \bibinfo{author}{\bibfnamefont{F.}~\bibnamefont{Schedin}},
  \bibinfo{author}{\bibfnamefont{T.~J.} \bibnamefont{Booth}},
  \bibinfo{author}{\bibfnamefont{V.~V.} \bibnamefont{Khotkevich}},
  \bibinfo{author}{\bibfnamefont{S.}~\bibnamefont{Morozov}}, \bibnamefont{and}
  \bibinfo{author}{\bibfnamefont{A.}~\bibnamefont{Geim}},
  \bibinfo{journal}{Proc. Natl. Acad. Sci. (USA)}
  \textbf{\bibinfo{volume}{102}}, \bibinfo{pages}{10451}
  (\bibinfo{year}{2005}).

\bibitem[{ \citenamefont{Butler et~al.}(2013) \citenamefont{Butler, Hollen, Cao,
  Cui, Gupta, Guti{\'e}rrez, Heinz, Hong, Huang, Ismach et~al.}}]{2DACSNano}
\bibinfo{author}{\bibfnamefont{S.~Z.} \bibnamefont{Butler}},
  \bibinfo{author}{\bibfnamefont{S.~M.} \bibnamefont{Hollen}},
  \bibinfo{author}{\bibfnamefont{L.}~\bibnamefont{Cao}},
  \bibinfo{author}{\bibfnamefont{Y.}~\bibnamefont{Cui}},
  \bibinfo{author}{\bibfnamefont{J.~A.} \bibnamefont{Gupta}},
  \bibinfo{author}{\bibfnamefont{H.~R.} \bibnamefont{Guti{\'e}rrez}},
  \bibinfo{author}{\bibfnamefont{T.~F.} \bibnamefont{Heinz}},
  \bibinfo{author}{\bibfnamefont{S.~S.} \bibnamefont{Hong}},
  \bibinfo{author}{\bibfnamefont{J.}~\bibnamefont{Huang}},
  \bibinfo{author}{\bibfnamefont{A.~F.} \bibnamefont{Ismach}},
  \bibnamefont{et~al.}, \bibinfo{journal}{ACS Nano}
  \textbf{\bibinfo{volume}{7}}, \bibinfo{pages}{2898} (\bibinfo{year}{2013}).

\bibitem[{ \citenamefont{Bridgman}(1914)}]{Bridgman14}
\bibinfo{author}{\bibfnamefont{P.~W.} \bibnamefont{Bridgman}},
  \bibinfo{journal}{J. Am. Chem. Soc.} \textbf{\bibinfo{volume}{36}},
  \bibinfo{pages}{1344} (\bibinfo{year}{1914}),

\bibitem[{ \citenamefont{Keyes}(1953)}]{Keyes}
\bibinfo{author}{\bibfnamefont{R.~W.} \bibnamefont{Keyes}},
  \bibinfo{journal}{Phys. Rev.} \textbf{\bibinfo{volume}{92}},
  \bibinfo{pages}{580} (\bibinfo{year}{1953}).

\bibitem[{ \citenamefont{Ellis and Warschauer}(1962)}]{Ellis}
\bibinfo{author}{\bibfnamefont{R.~C.} \bibnamefont{Ellis}} \bibnamefont{and}
  \bibinfo{author}{\bibfnamefont{D.~M.} \bibnamefont{Warschauer}},
  \bibinfo{journal}{J. Electrochem. Soc.} \textbf{\bibinfo{volume}{109}},
  \bibinfo{pages}{C207} (\bibinfo{year}{1962}).

\bibitem[{ \citenamefont{Jamieson}(1963)}]{Jamieson}
\bibinfo{author}{\bibfnamefont{J.~C.} \bibnamefont{Jamieson}},
  \bibinfo{journal}{Science} \textbf{\bibinfo{volume}{139}},
  \bibinfo{pages}{1291} (\bibinfo{year}{1963}).

\bibitem[{ \citenamefont{Schiferl}(1979)}]{Schiferl}
\bibinfo{author}{\bibfnamefont{D.}~\bibnamefont{Schiferl}},
  \bibinfo{journal}{Phys. Rev. B} \textbf{\bibinfo{volume}{19}},
  \bibinfo{pages}{806} (\bibinfo{year}{1979}).

\bibitem[{ \citenamefont{Asahina et~al.}(1982) \citenamefont{Asahina, Shindo, and
  Morita}}]{Morita}
\bibinfo{author}{\bibfnamefont{H.}~\bibnamefont{Asahina}},
  \bibinfo{author}{\bibfnamefont{K.}~\bibnamefont{Shindo}}, \bibnamefont{and}
  \bibinfo{author}{\bibfnamefont{A.}~\bibnamefont{Morita}},
  \bibinfo{journal}{J. Phys. Soc. Jap.} \textbf{\bibinfo{volume}{51}},
  \bibinfo{pages}{1193} (\bibinfo{year}{1982}).

\bibitem[{ \citenamefont{Kawamura et~al.}(1984) \citenamefont{Kawamura,
  Shirotani, and Tachikawa}}]{SSC84}
\bibinfo{author}{\bibfnamefont{H.}~\bibnamefont{Kawamura}},
  \bibinfo{author}{\bibfnamefont{I.}~\bibnamefont{Shirotani}},
  \bibnamefont{and}
  \bibinfo{author}{\bibfnamefont{K.}~\bibnamefont{Tachikawa}},
  \bibinfo{journal}{Solid State Comm.} \textbf{\bibinfo{volume}{49}},
  \bibinfo{pages}{879} (\bibinfo{year}{1984}).

\bibitem[{ \citenamefont{Akahama et~al.}(2000) \citenamefont{Akahama, Kawamura,
  Carlson, {Le Bihan}, and H{\"a}usermann}}]{PRB2000}
\bibinfo{author}{\bibfnamefont{Y.}~\bibnamefont{Akahama}},
  \bibinfo{author}{\bibfnamefont{H.}~\bibnamefont{Kawamura}},
  \bibinfo{author}{\bibfnamefont{S.}~\bibnamefont{Carlson}},
  \bibinfo{author}{\bibfnamefont{T.}~\bibnamefont{{Le Bihan}}},
  \bibnamefont{and}
  \bibinfo{author}{\bibfnamefont{D.}~\bibnamefont{H{\"a}usermann}},
  \bibinfo{journal}{Phys. Rev. B} \textbf{\bibinfo{volume}{61}},
  \bibinfo{pages}{3139} (\bibinfo{year}{2000}).

\bibitem[{ \citenamefont{Monaco et~al.}(2003) \citenamefont{Monaco, Falconi,
  Crichton, and Mezouar}}]{PRL2003}
\bibinfo{author}{\bibfnamefont{G.}~\bibnamefont{Monaco}},
  \bibinfo{author}{\bibfnamefont{S.}~\bibnamefont{Falconi}},
  \bibinfo{author}{\bibfnamefont{W.~A.} \bibnamefont{Crichton}},
  \bibnamefont{and} \bibinfo{author}{\bibfnamefont{M.}~\bibnamefont{Mezouar}},
  \bibinfo{journal}{Phys. Rev. Lett.} \textbf{\bibinfo{volume}{90}},
  \bibinfo{pages}{255701} (\bibinfo{year}{2003}).

\bibitem[{ \citenamefont{Liu et~al.}(2014{\natexlab{a}}) \citenamefont{Liu, Neal,
  Zhu, Luo, Xu, Tom{\'a}nek, and Ye}}]{Peide}
\bibinfo{author}{\bibfnamefont{H.}~\bibnamefont{Liu}},
  \bibinfo{author}{\bibfnamefont{A.~T.} \bibnamefont{Neal}},
  \bibinfo{author}{\bibfnamefont{Z.}~\bibnamefont{Zhu}},
  \bibinfo{author}{\bibfnamefont{Z.}~\bibnamefont{Luo}},
  \bibinfo{author}{\bibfnamefont{X.}~\bibnamefont{Xu}},
  \bibinfo{author}{\bibfnamefont{D.}~\bibnamefont{Tom{\'a}nek}},
  \bibnamefont{and} \bibinfo{author}{\bibfnamefont{P.~D.} \bibnamefont{Ye}},
  \bibinfo{journal}{ACS Nano} \textbf{\bibinfo{volume}{8}},
  \bibinfo{pages}{4033} (\bibinfo{year}{2014}{\natexlab{a}}).

\bibitem[{ \citenamefont{Li et~al.}(2014) \citenamefont{Li, Yu, Ye, Ge, Ou, Wu,
  Feng, Chen, and Zhang}}]{Yuanbo}
\bibinfo{author}{\bibfnamefont{L.}~\bibnamefont{Li}},
  \bibinfo{author}{\bibfnamefont{Y.}~\bibnamefont{Yu}},
  \bibinfo{author}{\bibfnamefont{G.~J.} \bibnamefont{Ye}},
  \bibinfo{author}{\bibfnamefont{Q.}~\bibnamefont{Ge}},
  \bibinfo{author}{\bibfnamefont{X.}~\bibnamefont{Ou}},
  \bibinfo{author}{\bibfnamefont{H.}~\bibnamefont{Wu}},
  \bibinfo{author}{\bibfnamefont{D.}~\bibnamefont{Feng}},
  \bibinfo{author}{\bibfnamefont{X.~H.} \bibnamefont{Chen}}, \bibnamefont{and}
  \bibinfo{author}{\bibfnamefont{Y.}~\bibnamefont{Zhang}},
  \bibinfo{journal}{Nature Nanotech.} \textbf{\bibinfo{volume}{9}},
  \bibinfo{pages}{372} (\bibinfo{year}{2014}).

\bibitem[{ \citenamefont{Koenig et~al.}(2014) \citenamefont{Koenig, Doganov,
  Schmidt, {Castro Neto}, and {{\"O}zyilmaz}}}]{Koenig}
\bibinfo{author}{\bibfnamefont{S.~P.} \bibnamefont{Koenig}},
  \bibinfo{author}{\bibfnamefont{R.~A.} \bibnamefont{Doganov}},
  \bibinfo{author}{\bibfnamefont{H.}~\bibnamefont{Schmidt}},
  \bibinfo{author}{\bibfnamefont{A.~H.} \bibnamefont{{Castro Neto}}},
  \bibnamefont{and}
  \bibinfo{author}{\bibfnamefont{B.}~\bibnamefont{{{\"O}zyilmaz}}},
  \bibinfo{journal}{Appl. Phys. Lett.} \textbf{\bibinfo{volume}{104}},
  \bibinfo{pages}{103106} (\bibinfo{year}{2014}).

\bibitem[{ \citenamefont{Buscema et~al.}(2014) \citenamefont{Buscema,
  Groenendijk, Blanter, Steele, {van der Zant}, and
  Castellanos-Gomez}}]{Buscema}
\bibinfo{author}{\bibfnamefont{M.}~\bibnamefont{Buscema}},
  \bibinfo{author}{\bibfnamefont{D.~J.} \bibnamefont{Groenendijk}},
  \bibinfo{author}{\bibfnamefont{S.~I.} \bibnamefont{Blanter}},
  \bibinfo{author}{\bibfnamefont{G.~A.} \bibnamefont{Steele}},
  \bibinfo{author}{\bibfnamefont{H.~S.~J.} \bibnamefont{{van der Zant}}},
  \bibnamefont{and}
  \bibinfo{author}{\bibfnamefont{A.}~\bibnamefont{Castellanos-Gomez}},
  \bibinfo{journal}{Nano Lett.} \textbf{\bibinfo{volume}{14}},
  \bibinfo{pages}{3347} (\bibinfo{year}{2014}).

\bibitem[{ \citenamefont{Tran et~al.}(2014) \citenamefont{Tran, Soklaski, Liang,
  and Yang}}]{ani1}
\bibinfo{author}{\bibfnamefont{V.}~\bibnamefont{Tran}},
  \bibinfo{author}{\bibfnamefont{R.}~\bibnamefont{Soklaski}},
  \bibinfo{author}{\bibfnamefont{Y.}~\bibnamefont{Liang}}, \bibnamefont{and}
  \bibinfo{author}{\bibfnamefont{L.}~\bibnamefont{Yang}},
  \bibinfo{journal}{Phys. Rev. B} \textbf{\bibinfo{volume}{89}},
  \bibinfo{pages}{235319} (\bibinfo{year}{2014}).

\bibitem[{ \citenamefont{Churchill and Jarillo-Herrero}(2014)}]{Jarillo}
\bibinfo{author}{\bibfnamefont{H.~O.~H.} \bibnamefont{Churchill}}
  \bibnamefont{and}
  \bibinfo{author}{\bibfnamefont{P.}~\bibnamefont{Jarillo-Herrero}},
  \bibinfo{journal}{Nature Nanotech.} \textbf{\bibinfo{volume}{9}},
  \bibinfo{pages}{330} (\bibinfo{year}{2014}).

\bibitem[{ \citenamefont{Fei and Yang}(2014)}]{strain1}
\bibinfo{author}{\bibfnamefont{R.}~\bibnamefont{Fei}} \bibnamefont{and}
  \bibinfo{author}{\bibfnamefont{L.}~\bibnamefont{Yang}},
  \bibinfo{journal}{Nano Lett.} \textbf{\bibinfo{volume}{14}},
  \bibinfo{pages}{2884} (\bibinfo{year}{2014}).

\bibitem[{ \citenamefont{Rodin et~al.}(2014) \citenamefont{Rodin, Carvalho, and
  {Castro Neto}}}]{strain2}
\bibinfo{author}{\bibfnamefont{A.~S.} \bibnamefont{Rodin}},
  \bibinfo{author}{\bibfnamefont{A.}~\bibnamefont{Carvalho}}, \bibnamefont{and}
  \bibinfo{author}{\bibfnamefont{A.~H.} \bibnamefont{{Castro Neto}}},
  \bibinfo{journal}{Phys. Rev. Lett.} \textbf{\bibinfo{volume}{112}},
  \bibinfo{pages}{176801} (\bibinfo{year}{2014}).

\bibitem[{ \citenamefont{Rudenko and Katsnelson}(2014)}]{strain3}
\bibinfo{author}{\bibfnamefont{A.~N.} \bibnamefont{Rudenko}} \bibnamefont{and}
  \bibinfo{author}{\bibfnamefont{M.~I.} \bibnamefont{Katsnelson}},
  \bibinfo{journal}{Phys. Rev. B} \textbf{\bibinfo{volume}{89}},
  \bibinfo{pages}{201408} (\bibinfo{year}{2014}).

\bibitem[{ \citenamefont{Wei and Peng}(2014)}]{strain4}
\bibinfo{author}{\bibfnamefont{Q.}~\bibnamefont{Wei}} \bibnamefont{and}
  \bibinfo{author}{\bibfnamefont{X.}~\bibnamefont{Peng}},
  \bibinfo{journal}{Appl. Phys. Lett.} \textbf{\bibinfo{volume}{104}},
  \bibinfo{pages}{251915} (\bibinfo{year}{2014}).

\bibitem[{ \citenamefont{Jiang and Park}(2014)}]{strain5}
\bibinfo{author}{\bibfnamefont{J.-W.} \bibnamefont{Jiang}} \bibnamefont{and}
  \bibinfo{author}{\bibfnamefont{H.~S.} \bibnamefont{Park}},
  \bibinfo{journal}{Nature Comm.} \textbf{\bibinfo{volume}{5}},
  \bibinfo{pages}{4727} (\bibinfo{year}{2014}).

\bibitem[{ \citenamefont{Peng et~al.}(2014) \citenamefont{Peng, Wei, and
  Copple}}]{strain6}
\bibinfo{author}{\bibfnamefont{X.}~\bibnamefont{Peng}},
  \bibinfo{author}{\bibfnamefont{Q.}~\bibnamefont{Wei}}, \bibnamefont{and}
  \bibinfo{author}{\bibfnamefont{A.}~\bibnamefont{Copple}},
  \bibinfo{journal}{Phys. Rev. B} \textbf{\bibinfo{volume}{90}},
  \bibinfo{pages}{085402} (\bibinfo{year}{2014}).

\bibitem[{ \citenamefont{Zhu and Tom{\'a}nek}(2014)}]{Tomanek1}
\bibinfo{author}{\bibfnamefont{Z.}~\bibnamefont{Zhu}} \bibnamefont{and}
  \bibinfo{author}{\bibfnamefont{D.}~\bibnamefont{Tom{\'a}nek}},
  \bibinfo{journal}{Phys. Rev. Lett.} \textbf{\bibinfo{volume}{112}},
  \bibinfo{pages}{176802} (\bibinfo{year}{2014}).

\bibitem[{ \citenamefont{Guan et~al.}(2014{\natexlab{a}}) \citenamefont{Guan,
  Zhu, and Tom{\'a}nek}}]{Tomanek2}
\bibinfo{author}{\bibfnamefont{J.}~\bibnamefont{Guan}},
  \bibinfo{author}{\bibfnamefont{Z.}~\bibnamefont{Zhu}}, \bibnamefont{and}
  \bibinfo{author}{\bibfnamefont{D.}~\bibnamefont{Tom{\'a}nek}},
  \bibinfo{journal}{Phys. Rev. Lett.} \textbf{\bibinfo{volume}{113}},
  \bibinfo{pages}{046804} (\bibinfo{year}{2014}{\natexlab{a}}).

\bibitem[{ \citenamefont{Guan et~al.}(2014{\natexlab{b}}) \citenamefont{Guan,
  Zhu, and Tom\'anek}}]{Tiling}
\bibinfo{author}{\bibfnamefont{J.}~\bibnamefont{Guan}},
  \bibinfo{author}{\bibfnamefont{Z.}~\bibnamefont{Zhu}}, \bibnamefont{and}
  \bibinfo{author}{\bibfnamefont{D.}~\bibnamefont{Tom\'anek}},
  \bibinfo{journal}{ACS Nano} \textbf{\bibinfo{volume}{8}}, 
    \bibinfo{pages}{12763} (\bibinfo{year}{2014}{\natexlab{a}}). 

  
  \bibinfo{pages}{Article ASAP.
  10.1021/nn5059248} (\bibinfo{year}{2014}{\natexlab{b}}).

\bibitem[{ \citenamefont{Guan et~al.}(2014{\natexlab{c}}) \citenamefont{Guan,
  Zhu, and Tom\'anek}}]{DT235}
\bibinfo{author}{\bibfnamefont{J.}~\bibnamefont{Guan}},
  \bibinfo{author}{\bibfnamefont{Z.}~\bibnamefont{Zhu}}, \bibnamefont{and}
  \bibinfo{author}{\bibfnamefont{D.}~\bibnamefont{Tom\'anek}},
  \bibinfo{journal}{Phys. Rev. Lett.} \textbf{\bibinfo{volume}{113}},
  \bibinfo{pages}{226801} (\bibinfo{year}{2014}{\natexlab{c}}).

\bibitem[{ \citenamefont{Bocker and Haser}(1995)}]{polydiversity0}
\bibinfo{author}{\bibfnamefont{S.}~\bibnamefont{Bocker}} \bibnamefont{and}
  \bibinfo{author}{\bibfnamefont{M.}~\bibnamefont{Haser}}, \bibinfo{journal}{Z.
  Anorg. Allg. Chem.} \textbf{\bibinfo{volume}{621}}, \bibinfo{pages}{258}
  (\bibinfo{year}{1995}).

\bibitem[{ \citenamefont{Seifert and Hernandez}(2000)}]{polydiversity1}
\bibinfo{author}{\bibfnamefont{G.}~\bibnamefont{Seifert}} \bibnamefont{and}
  \bibinfo{author}{\bibfnamefont{E.}~\bibnamefont{Hernandez}},
  \bibinfo{journal}{Chem. Phys. Lett.} \textbf{\bibinfo{volume}{318}},
  \bibinfo{pages}{355} (\bibinfo{year}{2000}).

\bibitem[{ \citenamefont{Karttunen et~al.}(2008) \citenamefont{Karttunen,
  Linnolahti, and Pakkanen}}]{polydiversity2}
\bibinfo{author}{\bibfnamefont{A.~J.} \bibnamefont{Karttunen}},
  \bibinfo{author}{\bibfnamefont{M.}~\bibnamefont{Linnolahti}},
  \bibnamefont{and} \bibinfo{author}{\bibfnamefont{T.~A.}
  \bibnamefont{Pakkanen}}, \bibinfo{journal}{Chem. Phys. Phys. Chem.}
  \textbf{\bibinfo{volume}{9}}, \bibinfo{pages}{2550} (\bibinfo{year}{2008}).

\bibitem[{ \citenamefont{Boulfelfel et~al.}(2012) \citenamefont{Boulfelfel,
  Seifert, Grin, and Leoni}}]{polydiversity3}
\bibinfo{author}{\bibfnamefont{S.~E.} \bibnamefont{Boulfelfel}},
  \bibinfo{author}{\bibfnamefont{G.}~\bibnamefont{Seifert}},
  \bibinfo{author}{\bibfnamefont{Y.}~\bibnamefont{Grin}}, \bibnamefont{and}
  \bibinfo{author}{\bibfnamefont{S.}~\bibnamefont{Leoni}},
  \bibinfo{journal}{Phys. Rev. B} \textbf{\bibinfo{volume}{85}},
  \bibinfo{pages}{014110} (\bibinfo{year}{2012}).

\bibitem[{ \citenamefont{Han et~al.}(2014) \citenamefont{Han, Stewart, Shevlin,
  Catlow, and Guo}}]{NLribbons}
\bibinfo{author}{\bibfnamefont{X.}~\bibnamefont{Han}},
  \bibinfo{author}{\bibfnamefont{H.~M.} \bibnamefont{Stewart}},
  \bibinfo{author}{\bibfnamefont{S.~A.} \bibnamefont{Shevlin}},
  \bibinfo{author}{\bibfnamefont{C.~R.~A.} \bibnamefont{Catlow}},
  \bibnamefont{and} \bibinfo{author}{\bibfnamefont{Z.~X.} \bibnamefont{Guo}},
  \bibinfo{journal}{Nano Lett.} \textbf{\bibinfo{volume}{14}},
  \bibinfo{pages}{4607} (\bibinfo{year}{2014}).

\bibitem[{ \citenamefont{Castellanos-Gomez
  et~al.}(2014) \citenamefont{Castellanos-Gomez, Vicarelli, Prada, Island,
  Narasimha-Acharya, Blanter, Groenendijk, Buscema, Steele, Alvarez
  et~al.}}]{ambient}
\bibinfo{author}{\bibfnamefont{A.}~\bibnamefont{Castellanos-Gomez}},
  \bibinfo{author}{\bibfnamefont{L.}~\bibnamefont{Vicarelli}},
  \bibinfo{author}{\bibfnamefont{E.}~\bibnamefont{Prada}},
  \bibinfo{author}{\bibfnamefont{J.~O.} \bibnamefont{Island}},
  \bibinfo{author}{\bibfnamefont{K.~L.} \bibnamefont{Narasimha-Acharya}},
  \bibinfo{author}{\bibfnamefont{S.~I.} \bibnamefont{Blanter}},
  \bibinfo{author}{\bibfnamefont{D.~J.} \bibnamefont{Groenendijk}},
  \bibinfo{author}{\bibfnamefont{M.}~\bibnamefont{Buscema}},
  \bibinfo{author}{\bibfnamefont{G.~A.} \bibnamefont{Steele}},
  \bibinfo{author}{\bibfnamefont{J.~V.} \bibnamefont{Alvarez}},
  \bibnamefont{et~al.}, \bibinfo{journal}{2D Materials}
  \textbf{\bibinfo{volume}{1}}, \bibinfo{pages}{025001} (\bibinfo{year}{2014}).

\bibitem[{ \citenamefont{Wood et~al.}(2014) \citenamefont{Wood, Wells, Jariwala,
  Chen, Cho, Sangwan, Liu, Lauhon, Marks, and Hersam}}]{Hersam}
\bibinfo{author}{\bibfnamefont{J.~D.} \bibnamefont{Wood}},
  \bibinfo{author}{\bibfnamefont{S.~A.} \bibnamefont{Wells}},
  \bibinfo{author}{\bibfnamefont{D.}~\bibnamefont{Jariwala}},
  \bibinfo{author}{\bibfnamefont{K.-S.} \bibnamefont{Chen}},
  \bibinfo{author}{\bibfnamefont{E.}~\bibnamefont{Cho}},
  \bibinfo{author}{\bibfnamefont{V.~K.} \bibnamefont{Sangwan}},
  \bibinfo{author}{\bibfnamefont{X.}~\bibnamefont{Liu}},
  \bibinfo{author}{\bibfnamefont{L.~J.} \bibnamefont{Lauhon}},
  \bibinfo{author}{\bibfnamefont{T.~J.} \bibnamefont{Marks}}, \bibnamefont{and}
  \bibinfo{author}{\bibfnamefont{M.~C.} \bibnamefont{Hersam}},
  \bibinfo{journal}{Nano Lett.}  \textbf{\bibinfo{volume}{14}}, \bibinfo{pages}{6964} (\bibinfo{year}{2014}).
  
\bibitem[{ \citenamefont{Tom\'anek}(2014)}]{DT234}
\bibinfo{author}{\bibfnamefont{D.}~\bibnamefont{Tom\'anek}},
  \bibinfo{journal}{Mater. Express} \textbf{\bibinfo{volume}{4}},
  \bibinfo{pages}{545} (\bibinfo{year}{2014}).

\bibitem[{ \citenamefont{Shin et~al.}(2014) \citenamefont{Shin, Park, Ryu, Jung,
  and Kim}}]{Graphene_protector1}
\bibinfo{author}{\bibfnamefont{J.}~\bibnamefont{Shin}},
  \bibinfo{author}{\bibfnamefont{K.}~\bibnamefont{Park}},
  \bibinfo{author}{\bibfnamefont{W.-H.} \bibnamefont{Ryu}},
  \bibinfo{author}{\bibfnamefont{J.-W.} \bibnamefont{Jung}}, \bibnamefont{and}
  \bibinfo{author}{\bibfnamefont{I.-D.} \bibnamefont{Kim}},
  \bibinfo{journal}{Nanoscale} \textbf{\bibinfo{volume}{6}},
  \bibinfo{pages}{12718} (\bibinfo{year}{2014}).

\bibitem[{ \citenamefont{Chen et~al.}(2011) \citenamefont{Chen, Brown, Levendorf,
  Cai, Ju, Edgeworth, Li, Magnuson, Velamakanni, Piner
  et~al.}}]{Graphene_protector2}
\bibinfo{author}{\bibfnamefont{S.}~\bibnamefont{Chen}},
  \bibinfo{author}{\bibfnamefont{L.}~\bibnamefont{Brown}},
  \bibinfo{author}{\bibfnamefont{M.}~\bibnamefont{Levendorf}},
  \bibinfo{author}{\bibfnamefont{W.}~\bibnamefont{Cai}},
  \bibinfo{author}{\bibfnamefont{S.-Y.} \bibnamefont{Ju}},
  \bibinfo{author}{\bibfnamefont{J.}~\bibnamefont{Edgeworth}},
  \bibinfo{author}{\bibfnamefont{X.}~\bibnamefont{Li}},
  \bibinfo{author}{\bibfnamefont{C.~W.} \bibnamefont{Magnuson}},
  \bibinfo{author}{\bibfnamefont{A.}~\bibnamefont{Velamakanni}},
  \bibinfo{author}{\bibfnamefont{R.~D.} \bibnamefont{Piner}},
  \bibnamefont{et~al.}, \bibinfo{journal}{ACS Nano}
  \textbf{\bibinfo{volume}{5}}, \bibinfo{pages}{1321} (\bibinfo{year}{2011}).

\bibitem[{ \citenamefont{Coan et~al.}(2013) \citenamefont{Coan, Barroso, Motz,
  Bolz\'{a}n, and Machado}}]{protective1}
\bibinfo{author}{\bibfnamefont{T.}~\bibnamefont{Coan}},
  \bibinfo{author}{\bibfnamefont{G.~S.} \bibnamefont{Barroso}},
  \bibinfo{author}{\bibfnamefont{G.}~\bibnamefont{Motz}},
  \bibinfo{author}{\bibfnamefont{A.}~\bibnamefont{Bolz\'{a}n}},
  \bibnamefont{and} \bibinfo{author}{\bibfnamefont{R.~A.~F.}
  \bibnamefont{Machado}}, \bibinfo{journal}{Mater. Res.}
  \textbf{\bibinfo{volume}{16}}, \bibinfo{pages}{1366} (\bibinfo{year}{2013}).

\bibitem[{ \citenamefont{Gillgren et~al.}(2014) \citenamefont{Gillgren,
  Wickramaratne, Shi, Espiritu, Yang, Hu, Wei, Liu, Mao, Watanabe
  et~al.}}]{Lau}
\bibinfo{author}{\bibfnamefont{N.}~\bibnamefont{Gillgren}},
  \bibinfo{author}{\bibfnamefont{D.}~\bibnamefont{Wickramaratne}},
  \bibinfo{author}{\bibfnamefont{Y.}~\bibnamefont{Shi}},
  \bibinfo{author}{\bibfnamefont{T.}~\bibnamefont{Espiritu}},
  \bibinfo{author}{\bibfnamefont{J.}~\bibnamefont{Yang}},
  \bibinfo{author}{\bibfnamefont{J.}~\bibnamefont{Hu}},
  \bibinfo{author}{\bibfnamefont{J.}~\bibnamefont{Wei}},
  \bibinfo{author}{\bibfnamefont{X.}~\bibnamefont{Liu}},
  \bibinfo{author}{\bibfnamefont{Z.}~\bibnamefont{Mao}},
  \bibinfo{author}{\bibfnamefont{K.}~\bibnamefont{Watanabe}},
  \bibnamefont{et~al.} \bibinfo{journal}{2D Materials}
  \textbf{\bibinfo{volume}{2}}, \bibinfo{pages}{011001} (\bibinfo{year}{2015}).
  
\bibitem[{ \citenamefont{Dean et~al.}(2010{\natexlab{a}}) \citenamefont{Dean,
  Young, Meric, Lee, Wang, Sorgenfrei, Watanabe, Taniguchi, Kim, Shepard
  et~al.}}]{hBN-mech1}
\bibinfo{author}{\bibfnamefont{C.~R.} \bibnamefont{Dean}},
  \bibinfo{author}{\bibfnamefont{A.~F.} \bibnamefont{Young}},
  \bibinfo{author}{\bibfnamefont{I.}~\bibnamefont{Meric}},
  \bibinfo{author}{\bibfnamefont{C.}~\bibnamefont{Lee}},
  \bibinfo{author}{\bibfnamefont{L.}~\bibnamefont{Wang}},
  \bibinfo{author}{\bibfnamefont{S.}~\bibnamefont{Sorgenfrei}},
  \bibinfo{author}{\bibfnamefont{K.}~\bibnamefont{Watanabe}},
  \bibinfo{author}{\bibfnamefont{T.}~\bibnamefont{Taniguchi}},
  \bibinfo{author}{\bibfnamefont{P.}~\bibnamefont{Kim}},
  \bibinfo{author}{\bibfnamefont{K.~L.} \bibnamefont{Shepard}},
  \bibnamefont{et~al.}, \bibinfo{journal}{Nat. Nanotechnol.}
  \textbf{\bibinfo{volume}{5}}, \bibinfo{pages}{722}
  (\bibinfo{year}{2010}{\natexlab{a}}).

\bibitem[{ \citenamefont{Pacil\'e et~al.}(2008) \citenamefont{Pacil\'e, Meyer,
  Girit, and Zettl}}]{hBN-mech2}
\bibinfo{author}{\bibfnamefont{D.}~\bibnamefont{Pacil\'e}},
  \bibinfo{author}{\bibfnamefont{J.~C.} \bibnamefont{Meyer}},
  \bibinfo{author}{\bibfnamefont{C.~O.} \bibnamefont{Girit}}, \bibnamefont{and}
  \bibinfo{author}{\bibfnamefont{A.}~\bibnamefont{Zettl}},
  \bibinfo{journal}{Appl. Phys. Lett.} \textbf{\bibinfo{volume}{92}},
  \bibinfo{pages}{133107} (\bibinfo{year}{2008}).

\bibitem[{ \citenamefont{Lee et~al.}(2010) \citenamefont{Lee, Li, Kalb, Liu,
  Berger, Carpick, and Hone}}]{hBN-mech3}
\bibinfo{author}{\bibfnamefont{C.~G.} \bibnamefont{Lee}},
  \bibinfo{author}{\bibfnamefont{Q.}~\bibnamefont{Li}},
  \bibinfo{author}{\bibfnamefont{W.}~\bibnamefont{Kalb}},
  \bibinfo{author}{\bibfnamefont{X.}~\bibnamefont{Liu}},
  \bibinfo{author}{\bibfnamefont{H.}~\bibnamefont{Berger}},
  \bibinfo{author}{\bibfnamefont{R.~W.} \bibnamefont{Carpick}},
  \bibnamefont{and} \bibinfo{author}{\bibfnamefont{J.}~\bibnamefont{Hone}},
  \bibinfo{journal}{Science} \textbf{\bibinfo{volume}{328}},
  \bibinfo{pages}{76} (\bibinfo{year}{2010}).

\bibitem[{ \citenamefont{Dean et~al.}(2010{\natexlab{b}}) \citenamefont{Dean,
  Young, Meric, Lee, Wang, Sorgenfrei, Watanabe, Taniguchi, Kim, Shepard
  et~al.}}]{hBN_subst1}
\bibinfo{author}{\bibfnamefont{C.~R.} \bibnamefont{Dean}},
  \bibinfo{author}{\bibfnamefont{A.~F.} \bibnamefont{Young}},
  \bibinfo{author}{\bibfnamefont{I.}~\bibnamefont{Meric}},
  \bibinfo{author}{\bibfnamefont{C.}~\bibnamefont{Lee}},
  \bibinfo{author}{\bibfnamefont{L.}~\bibnamefont{Wang}},
  \bibinfo{author}{\bibfnamefont{S.}~\bibnamefont{Sorgenfrei}},
  \bibinfo{author}{\bibfnamefont{K.}~\bibnamefont{Watanabe}},
  \bibinfo{author}{\bibfnamefont{T.}~\bibnamefont{Taniguchi}},
  \bibinfo{author}{\bibfnamefont{P.}~\bibnamefont{Kim}},
  \bibinfo{author}{\bibfnamefont{K.~L.} \bibnamefont{Shepard}},
  \bibnamefont{et~al.}, \bibinfo{journal}{Nat. Mater.}
  \textbf{\bibinfo{volume}{5}}, \bibinfo{pages}{722}
  (\bibinfo{year}{2010}{\natexlab{b}}).

\bibitem[{ \citenamefont{Kim et~al.}(2012) \citenamefont{Kim, Hsu, Jia, Kim, Shi,
  Hofmann, Nezich, Rodriguez-Nieva, Dresselhaus, Palacios et~al.}}]{hBN-CVD}
\bibinfo{author}{\bibfnamefont{K.~K.} \bibnamefont{Kim}},
  \bibinfo{author}{\bibfnamefont{A.}~\bibnamefont{Hsu}},
  \bibinfo{author}{\bibfnamefont{X.}~\bibnamefont{Jia}},
  \bibinfo{author}{\bibfnamefont{S.~M.} \bibnamefont{Kim}},
  \bibinfo{author}{\bibfnamefont{Y.}~\bibnamefont{Shi}},
  \bibinfo{author}{\bibfnamefont{M.}~\bibnamefont{Hofmann}},
  \bibinfo{author}{\bibfnamefont{D.}~\bibnamefont{Nezich}},
  \bibinfo{author}{\bibfnamefont{J.~F.} \bibnamefont{Rodriguez-Nieva}},
  \bibinfo{author}{\bibfnamefont{M.}~\bibnamefont{Dresselhaus}},
  \bibinfo{author}{\bibfnamefont{T.}~\bibnamefont{Palacios}},
  \bibnamefont{et~al.}, \bibinfo{journal}{Nano Letters}
  \textbf{\bibinfo{volume}{12}}, \bibinfo{pages}{161} (\bibinfo{year}{2012}).

\bibitem[{ \citenamefont{Watanabe et~al.}(2004) \citenamefont{Watanabe,
  Taniguchi, and Kanda}}]{hBN_expgap}
\bibinfo{author}{\bibfnamefont{K.}~\bibnamefont{Watanabe}},
  \bibinfo{author}{\bibfnamefont{T.}~\bibnamefont{Taniguchi}},
  \bibnamefont{and} \bibinfo{author}{\bibfnamefont{H.}~\bibnamefont{Kanda}},
  \bibinfo{journal}{Nat. Mater} \textbf{\bibinfo{volume}{3}},
  \bibinfo{pages}{404} (\bibinfo{year}{2004}).

\bibitem[{ \citenamefont{He et~al.}(2010) \citenamefont{He, Koepke,
  Barraza-Lopez, and Lyding}}]{NLus}
\bibinfo{author}{\bibfnamefont{K.~T.} \bibnamefont{He}},
  \bibinfo{author}{\bibfnamefont{J.~C.} \bibnamefont{Koepke}},
  \bibinfo{author}{\bibfnamefont{S.}~\bibnamefont{Barraza-Lopez}},
  \bibnamefont{and} \bibinfo{author}{\bibfnamefont{J.~W.}
  \bibnamefont{Lyding}}, \bibinfo{journal}{Nano Lett.}
  \textbf{\bibinfo{volume}{10}}, \bibinfo{pages}{3446} (\bibinfo{year}{2010}).

\bibitem[{ \citenamefont{Liu et~al.}(2014{\natexlab{b}}) \citenamefont{Liu, Xu,
  Zhang, Penev, and Yakobson}}]{JakobsonDefects}
\bibinfo{author}{\bibfnamefont{Y.}~\bibnamefont{Liu}},
  \bibinfo{author}{\bibfnamefont{F.}~\bibnamefont{Xu}},
  \bibinfo{author}{\bibfnamefont{Z.}~\bibnamefont{Zhang}},
  \bibinfo{author}{\bibfnamefont{E.~S.} \bibnamefont{Penev}}, \bibnamefont{and}
  \bibinfo{author}{\bibfnamefont{B.~I.} \bibnamefont{Yakobson}},
  \bibinfo{journal}{Nano Lett.} p. \bibinfo{pages}{Article ASAP.
  10.1021/nl5021393} (\bibinfo{year}{2014}{\natexlab{b}}).

\bibitem[{ \citenamefont{H{\"o}ltzl et~al.}(2010) \citenamefont{H{\"o}ltzl,
  Veszpr{\'e}mi, and Nguyen}}]{p2010}
\bibinfo{author}{\bibfnamefont{T.}~\bibnamefont{H{\"o}ltzl}},
  \bibinfo{author}{\bibfnamefont{T.}~\bibnamefont{Veszpr{\'e}mi}},
  \bibnamefont{and} \bibinfo{author}{\bibfnamefont{M.~T.}
  \bibnamefont{Nguyen}}, \bibinfo{journal}{C. R. Chemie}
  \textbf{\bibinfo{volume}{13}}, \bibinfo{pages}{1173} (\bibinfo{year}{2010}).

\bibitem[{ \citenamefont{Artacho et~al.}(2008) \citenamefont{Artacho, Anglada,
  Dieguez, Gale, Garc\'{\i}a, Junquera, Martin, Ordej\'on, Pruneda,
  S\'anchez-Portal et~al.}}]{SIESTA}
\bibinfo{author}{\bibfnamefont{E.}~\bibnamefont{Artacho}},
  \bibinfo{author}{\bibfnamefont{E.}~\bibnamefont{Anglada}},
  \bibinfo{author}{\bibfnamefont{O.}~\bibnamefont{Dieguez}},
  \bibinfo{author}{\bibfnamefont{J.~D.} \bibnamefont{Gale}},
  \bibinfo{author}{\bibfnamefont{A.}~\bibnamefont{Garc\'{\i}a}},
  \bibinfo{author}{\bibfnamefont{J.}~\bibnamefont{Junquera}},
  \bibinfo{author}{\bibfnamefont{R.~M.} \bibnamefont{Martin}},
  \bibinfo{author}{\bibfnamefont{P.}~\bibnamefont{Ordej\'on}},
  \bibinfo{author}{\bibfnamefont{J.~M.} \bibnamefont{Pruneda}},
  \bibinfo{author}{\bibfnamefont{D.}~\bibnamefont{S\'anchez-Portal}},
  \bibnamefont{et~al.}, \bibinfo{journal}{J. Phys. Cond. Mat.}
  \textbf{\bibinfo{volume}{20}}, \bibinfo{pages}{064208}
  (\bibinfo{year}{2008}).

\bibitem[{ \citenamefont{Klimes et~al.}(2010) \citenamefont{Klimes, Bowler, and
  Michaelides}}]{KBM}
\bibinfo{author}{\bibfnamefont{J.}~\bibnamefont{Klimes}},
  \bibinfo{author}{\bibfnamefont{D.~R.} \bibnamefont{Bowler}},
  \bibnamefont{and}
  \bibinfo{author}{\bibfnamefont{A.}~\bibnamefont{Michaelides}},
  \bibinfo{journal}{J. Phys. Condens. Matter} \textbf{\bibinfo{volume}{22}},
  \bibinfo{pages}{022201} (\bibinfo{year}{2010}).

\bibitem[{ \citenamefont{Troullier and Martins}(1991)}]{Troullier}
\bibinfo{author}{\bibfnamefont{N.}~\bibnamefont{Troullier}} \bibnamefont{and}
  \bibinfo{author}{\bibfnamefont{J.~L.} \bibnamefont{Martins}},
  \bibinfo{journal}{Phys. Rev. B} \textbf{\bibinfo{volume}{43}},
  \bibinfo{pages}{1993} (\bibinfo{year}{1991}).

\bibitem[{ \citenamefont{Du et~al.}(2010) \citenamefont{Du, Ouyang, Shi, and
  Lei}}]{vdW1}
\bibinfo{author}{\bibfnamefont{Y.}~\bibnamefont{Du}},
  \bibinfo{author}{\bibfnamefont{C.}~\bibnamefont{Ouyang}},
  \bibinfo{author}{\bibfnamefont{S.}~\bibnamefont{Shi}}, \bibnamefont{and}
  \bibinfo{author}{\bibfnamefont{M.}~\bibnamefont{Lei}}, \bibinfo{journal}{J.
  Appl. Phys.} \textbf{\bibinfo{volume}{107}}, \bibinfo{pages}{093718}
  (\bibinfo{year}{2010}).

\bibitem[{ \citenamefont{Prytz and Flage-Larsen}(2010)}]{vdW2}
\bibinfo{author}{\bibfnamefont{{\O}.}~\bibnamefont{Prytz}} \bibnamefont{and}
  \bibinfo{author}{\bibfnamefont{E.}~\bibnamefont{Flage-Larsen}},
  \bibinfo{journal}{J. Phys.: Condens. Matter} \textbf{\bibinfo{volume}{22}},
  \bibinfo{pages}{015502} (\bibinfo{year}{2010}).

\bibitem[{ \citenamefont{Appalakondaiah
  et~al.}(2012) \citenamefont{Appalakondaiah, Vaitheeswaran, Leb{\'e}gue,
  Christensen, and Svane}}]{vdW3}
\bibinfo{author}{\bibfnamefont{S.}~\bibnamefont{Appalakondaiah}},
  \bibinfo{author}{\bibfnamefont{G.}~\bibnamefont{Vaitheeswaran}},
  \bibinfo{author}{\bibfnamefont{S.}~\bibnamefont{Leb{\'e}gue}},
  \bibinfo{author}{\bibfnamefont{N.~E.} \bibnamefont{Christensen}},
  \bibnamefont{and} \bibinfo{author}{\bibfnamefont{A.}~\bibnamefont{Svane}},
  \bibinfo{journal}{Phys. Rev. B} \textbf{\bibinfo{volume}{86}},
  \bibinfo{pages}{035105} (\bibinfo{year}{2012}).

\bibitem[{ \citenamefont{Tayran et~al.}(2013) \citenamefont{Tayran, Zhu, Baldoni,
  Selli, Seifert, and Tom\'anek}}]{DT220}
\bibinfo{author}{\bibfnamefont{C.}~\bibnamefont{Tayran}},
  \bibinfo{author}{\bibfnamefont{Z.}~\bibnamefont{Zhu}},
  \bibinfo{author}{\bibfnamefont{M.}~\bibnamefont{Baldoni}},
  \bibinfo{author}{\bibfnamefont{D.}~\bibnamefont{Selli}},
  \bibinfo{author}{\bibfnamefont{G.}~\bibnamefont{Seifert}}, \bibnamefont{and}
  \bibinfo{author}{\bibfnamefont{D.}~\bibnamefont{Tom\'anek}},
  \bibinfo{journal}{Phys. Rev. Lett.} \textbf{\bibinfo{volume}{110}},
  \bibinfo{pages}{176805} (\bibinfo{year}{2013}).

\bibitem[{ \citenamefont{Fujihisa et~al.}(2007) \citenamefont{Fujihisa, Akahama,
  Kawamura, Ohishi, Gotoh, Yamawaki, Sakashita, Takeya, and Honda}}]{PRLpiv}
\bibinfo{author}{\bibfnamefont{H.}~\bibnamefont{Fujihisa}},
  \bibinfo{author}{\bibfnamefont{Y.}~\bibnamefont{Akahama}},
  \bibinfo{author}{\bibfnamefont{H.}~\bibnamefont{Kawamura}},
  \bibinfo{author}{\bibfnamefont{Y.}~\bibnamefont{Ohishi}},
  \bibinfo{author}{\bibfnamefont{Y.}~\bibnamefont{Gotoh}},
  \bibinfo{author}{\bibfnamefont{H.}~\bibnamefont{Yamawaki}},
  \bibinfo{author}{\bibfnamefont{M.}~\bibnamefont{Sakashita}},
  \bibinfo{author}{\bibfnamefont{S.}~\bibnamefont{Takeya}}, \bibnamefont{and}
  \bibinfo{author}{\bibfnamefont{K.}~\bibnamefont{Honda}},
  \bibinfo{journal}{Phys. Rev. Lett.} \textbf{\bibinfo{volume}{98}},
  \bibinfo{pages}{175501} (\bibinfo{year}{2007}).


\bibitem{jpcm}
\bibinfo{author}{\bibfnamefont{N.}~\bibnamefont{Ooi}},
  \bibinfo{author}{\bibfnamefont{A.} \bibnamefont{Rairkar}},
  \bibinfo{author}{\bibfnamefont{L.} \bibnamefont{Lindsley}},
  \bibnamefont{and} \bibinfo{author}{\bibfnamefont{J.~B.}~\bibnamefont{Adams}},
  \bibinfo{journal}{J. Phys.: Condens. Matter} \textbf{\bibinfo{volume}{18}},
  \bibinfo{pages}{97} (\bibinfo{year}{2006}).

\bibitem[{ \citenamefont{Barraza-Lopez et~al.}(2006) \citenamefont{Barraza-Lopez,
  Albrecht, Romero, and Hess}}]{JAP2006}
\bibinfo{author}{\bibfnamefont{S.}~\bibnamefont{Barraza-Lopez}},
  \bibinfo{author}{\bibfnamefont{P.~M.} \bibnamefont{Albrecht}},
  \bibinfo{author}{\bibfnamefont{N.~A.} \bibnamefont{Romero}},
  \bibnamefont{and} \bibinfo{author}{\bibfnamefont{K.}~\bibnamefont{Hess}},
  \bibinfo{journal}{J. Appl. Phys.} \textbf{\bibinfo{volume}{100}},
  \bibinfo{pages}{124304} (\bibinfo{year}{2006}).

\bibitem[{ \citenamefont{Tomanek and Louie}(1988)}]{DT029}
\bibinfo{author}{\bibfnamefont{D.}~\bibnamefont{Tomanek}} \bibnamefont{and}
  \bibinfo{author}{\bibfnamefont{S.~G.} \bibnamefont{Louie}},
  \bibinfo{journal}{Phys. Rev. B} \textbf{\bibinfo{volume}{37}},
  \bibinfo{pages}{8327} (\bibinfo{year}{1988}).

\bibitem[{ \citenamefont{Selloni et~al.}(1985) \citenamefont{Selloni, Carnevali,
  Tosatti, and Chen}}]{STMSelloni85}
\bibinfo{author}{\bibfnamefont{A.}~\bibnamefont{Selloni}},
  \bibinfo{author}{\bibfnamefont{P.}~\bibnamefont{Carnevali}},
  \bibinfo{author}{\bibfnamefont{E.}~\bibnamefont{Tosatti}}, \bibnamefont{and}
  \bibinfo{author}{\bibfnamefont{C.}~\bibnamefont{Chen}},
  \bibinfo{journal}{Phys. Rev. B} \textbf{\bibinfo{volume}{31}},
  \bibinfo{pages}{2602} (\bibinfo{year}{1985}).

\bibitem[{ \citenamefont{Tersoff and Hamann}(1983)}]{STMPRL83}
\bibinfo{author}{\bibfnamefont{J.}~\bibnamefont{Tersoff}} \bibnamefont{and}
  \bibinfo{author}{\bibfnamefont{D.}~\bibnamefont{Hamann}},
  \bibinfo{journal}{Phys. Rev. Lett.} \textbf{\bibinfo{volume}{50}},
  \bibinfo{pages}{1998} (\bibinfo{year}{1983}).

\bibitem[{ \citenamefont{Tersoff and Hamann}(1985)}]{STMPRB85}
\bibinfo{author}{\bibfnamefont{J.}~\bibnamefont{Tersoff}} \bibnamefont{and}
  \bibinfo{author}{\bibfnamefont{D.}~\bibnamefont{Hamann}},
  \bibinfo{journal}{Phys. Rev. B} \textbf{\bibinfo{volume}{31}},
  \bibinfo{pages}{805} (\bibinfo{year}{1985}).

\end{thebibliography}
\end{document}